\documentclass{WileyChemistry-template}

\usepackage{subcaption}
\usepackage{stfloats}
\usepackage{xr}

\makeatletter
\newcommand*{\addFileDependency}[1]{
\typeout{(#1)}
\@addtofilelist{#1}
\IfFileExists{#1}{}{\typeout{No file #1.}}
}\makeatother

\newcommand*{\myexternaldocument}[1]{%
\externaldocument{#1}%
\addFileDependency{#1.tex}%
\addFileDependency{#1.aux}%
}

\myexternaldocument{main_supp}

\title{\raggedright Challenges in Measuring Transport Parameters of Carbonate-based Electrolytes}

\author{
\begin{minipage}{\textwidth}
	Lukas Lehnert,\textsuperscript{[a, b]} Maryam Nojabaee*,\textsuperscript{[b]} Arnulf Latz,\textsuperscript{[a, b, c]}  Birger Horstmann*\textsuperscript{[a, b, c]}   
\end{minipage}
}

\newcommand{\affiliation}{
\begin{itemize}

\item[{[a]}] L. Lehnert, Prof. Dr. A. Latz, Prof. Dr. B. Horstmann*\\
Helmholtz Institute Ulm\\
Helmholtzstraße 11, 89081 Ulm, Germany\\
E-mail: birger.horstmann@dlr.de

\item[{[b]}] L. Lehnert, Dr.~M. Nojabaee*, Prof.~Dr.~A. Latz, Prof.~Dr.~B.~Horstmann\\
German Aerospace Center\\
Pfaffenwaldring 38, 70569 Stuttgart, Germany\\
E-mail: maryam.nojabaee@dlr.de

\item[{[c]}] Prof. Dr. A. Latz, Prof. Dr. B. Horstmann\\
University of Ulm\\
Albert-Einstein-Allee 47, 89081 Ulm, Germany

\end{itemize}
}

\renewcommand{\dedication}{
}

\renewcommand{\abstract}{Numerical simulations are a powerful tool for the development and improvement of Li-ion batteries. Modeling the mass transport of the involved electrolytic solutions requires precise determination of the corresponding electrolyte parameters. In this work, we attempt to measure the conductivity, the diffusion coefficient, the transference number and the thermodynamic factor for a system of 0.5\,M LiPF$_6$ dissolved in a blend of ethylene carbonate and ethyl methyl carbonate (EC:EMC, 3:7 weight) at 20\,\textdegree C and 50\,\textdegree C. Applying galvanostatic polarization experiments to symmetrical metal Li | electrolyte + separator | Li metal cells reveals, however, a potential response qualitatively deviating from theoretical expectations. Impeded diffusion processes indicate the presence of additional, undesired porous structures on the Li electrodes, preventing a reliable evaluation of the electrolyte parameters. To spectrally resolve the diffusive processes, we conduct very-low-frequency impedance spectroscopy. The impedance in fact exhibits multiple interfering diffusive features. In our measurements, an explicit identification of the impedance for the sole diffusion through the separator is however not feasible. Therefore, the authors doubt that polarizing experiments using Li metal electrodes yield accurate parameters for electrolytes.
}

\newcommand{\keywords}{
        challenges \textbullet\ 
        carbonate-based electrolytes \textbullet\ 
        electrochemical measurements \textbullet\
	electrolyte parameters \textbullet\ 
	LiPF$_6$ in EC:EMC (3:7, weight)
}

\begin{document}
\emergencystretch 2em

\twocolumn[\vspace{-1.5cm}\maketitle\vspace{-1cm}
	\textit{\dedication}\vspace{0.4cm}]
\small{\begin{shaded}
		\noindent\abstract
	\end{shaded}
}

\begin{figure} [!b]
\begin{minipage}[t]{\columnwidth}{\rule{\columnwidth}{1pt}\footnotesize{\textsf{\affiliation}}}\end{minipage}
\end{figure}




\section{Introduction}
\label{introduction}

While Li-ion batteries have become prevalent in most markets for rechargeable batteries operating in portable devices and electric vehicles, research is still ongoing. To further improve and predict the performance of the cells, often numerical simulation tools are applied. The accuracy of these calculations depends on the precise characterization of the involved materials. Modeling the isobaric and isothermal mass transfer phenomena of $n$-component electrolytic solutions using the concentrated electrolyte theory from Latz et al. and Schammer et al. requires the determination of $(n-2)(n-1)/2$ unknown thermodynamic and $n(n-1)/2$ transport properties.\cite{Latz2011, Latz2015, Schammer2021} Thus, for electrolytes containing $n=3$ components four parameters have to be determined: the thermodynamic factor $TDF\left(c\right)$, the conductivity $\kappa\left(c\right)$, the salt diffusion coefficient $D_\pm\left(c\right)$ and the transference number $t_+\left(c\right)$.\\
The conductivity is readily accessible by measuring the high-frequency resistance of the bulk electrolyte with a calibrated measuring device, as for instance done in Refs.~\mbox{[\!\!\!\citenum{Landesfeind2019, Hou2020, Bergstrom2021}]}. However, the determination of $D_\pm\left(c\right)$, $t_+\left(c\right)$ and $TDF\left(c\right)$ remains sophisticated.\\
Nuclear magnetic resonance spectroscopy (NMR) measures the interaction of the examined species in a static magnetic field with a pulsed orthogonal oscillating high-frequency magnetic field. Introducing pulsed magnetic field gradients (PFGs) between the excitation pulse and data acquisition allows measuring the self-diffusion coefficients of the different electrolyte species. However, these coefficients translate into $t_+\left(c\right)$ only for highly diluted systems.\cite{Pesko2017, Buss2017} Using pulsed gradient spin echo (PGSE) experiments Klett and Sethurajan image the Li-ion concentration gradients over time while applying a constant current to the sample cell.\cite{Klett2012, Sethurajan2015} Combining the data with electrochemical transport models yields the salt diffusion coefficient and the transference number. In Electrophoretic NMR\cite{Fang2023,Rosenwinkel2019,Kilchert2023,Gouverneur2015} (eNMR) additional electric field pulses are applied. This drift enables direct determination of the transference number for concentrated electrolytes.
\begin{table*}[bt]
	\begin{center}
	\caption{Scheme of various authors determining the transference number using different techniques.} \label{pol_scheme}
		\begin{tabular}{cccc}	
\toprule		
Authors & Electrodes & Electrolyte & Method\\
\midrule
Nyman et al. \cite{Nyman2008}	& Li metal	& LiPF$_6$ in EC:EMC (3:7, weight)	& model fit to gal. pol. exp.\\
Lundgren et al. \cite{Lundgren2014}  & Li metal			& LiPF$_6$ in EC:DEC (1:1, weight)			& model fit to gal. pol. exp.\\
Lundgren et al. \cite{Lundgren2015}  & Li metal			& LiTFSI in ACN			& model fit to gal. pol. exp.\\
Wang et al. \cite{Wang2020}    & Li metal & LiPF$_6$ in EMC	& Hittorf \\
Wang et al. \cite{Wang2020} & Li metal & LiPF$_6$ in PC	& Hittorf\\
Hou et al. \cite{Hou2020}  & Li metal & LiPF$_6$ in PC & Hittorf\\
Val\o en and Reimers \cite{Valoen2005} & Li metal & LiPF$_6$ in PC:EC:DMC (10:27:63, vol.) & Hittorf\\
Landesfeind et al. \cite{Landesfeind2019} & Li metal & LiPF$_6$ in EC:DMC (1:1, weight) & current interrupt\\
Landesfeind et al. \cite{Landesfeind2019} & Li metal & LiPF$_6$ in EC:EMC (3:7, weight) & current interrupt\\
Landesfeind et al. \cite{Landesfeind2019} & Li metal & LiPF$_6$ in EMC:FEC (19:1, weight) & current interrupt\\
Ehrl et al. \cite{Ehrl2017} &  Li metal  & LiClO$_4$ in EC:DEC (1:1, weight)  & various pol. exp.\\
Bergstrom et al. \cite{Bergstrom2021}   & Li metal  & LiPF$_6$ in EC:EMC (3:7, weight)  & current ratio\\
\bottomrule	
	\end{tabular}
	\end{center}
\end{table*}\\
Harned and French introduced an electrochemical way to measure $D_\pm\left(c\right)$, the restricted diffusion method,\cite{Harned1945} extended by Newman and Chapman for concentrated electrolytes.\cite{Newman1973}
In this method, the time evolution of an initially generated concentration gradient is indirectly observed by measuring the induced concentration potential. While Doeff, Ferry and Ma measure the long-term relaxation of the potential after an excitation pulse,\cite{Doeff2000, Ferry1998, Ma1995} Hiller and Ehrl also determine the diffusion coefficient detecting the short-term potential subsequent to a linearly excited steady-state concentration gradient.\cite{Hiller2013, Ehrl2017a}\\
Concentration cells consist of two half-cells containing the same electrolyte with slightly different concentrations. Allowing negligibly small current densities between the half-cells enables measuring the potential between two immersed, active electrodes. This reveals convoluted information about the transference number and thermodynamic factor. To deconvolute these quantities, a second experiment is needed. For this,  Nyman et al. and Lundgren et al. fit the simulated relaxation of the open circuit potential to the corresponding data of galvanostatic polarization experiments, yielding the diffusion coefficient and the transference number.\cite{Nyman2008, Lundgren2014, Lundgren2015} While Hou et al., Wang et al., Val\o en and Reimers determine directly the transference number using a Hittorf cell \cite{Wang2020, Hou2020, Valoen2005}, Landesfeind et al. apply the current interrupt method by Ma and Newman.\cite{Landesfeind2019, Ma1995} Ehrl et al. compare various polarization methods.\cite{Ehrl2017} All of the listed polarization methods involve Li metal electrodes (see Table \ref{pol_scheme}). Many of them use similar electrolytes, containing blends of ethylene carbonate (EC), ethyl methyl carbonate (EMC), dimethyl carbonate (DMC), diethyl carbonate (DEC), and fluoroethylene carbonate (FEC).\\
In a recent paper, Bergstrom et al. investigated a system of lithium hexafluorophosphate (LiPF$_6$) dissolved in a blend of EC:EMC (3:7, weight) in Li-Li symmetric cells.\cite{Bergstrom2021} To determine the diffusion coefficient $D_\pm\left(c\right)$, the group examined the long-term relaxation of the potentials in restricted diffusion experiments. In contrast to the theory and the findings of Landesfeind et al., this study obtained varying values for $D_\pm\left(c\right)$ depending on the chosen time window used for fitting the potential. Therefore, the authors suggest restricting the evaluation of the potential data to the time interval during which the induced concentration gradient should mainly decay. This minimizes the influence of any non-diffusive phenomena.\\
Additionally, using the current ratio from polarization measurements, Bergstrom et al. found the measurement capturing not only the transport in the bulk electrolyte but also interfacial properties.\cite{Bergstrom2021, Balsara2015, Bruce1987, Evans1987} This leads in combination with a small ratio of electrolyte resistance to interfacial resistance to significant errors in the calculated transference numbers and thermodynamic factors. Therefore, Bergstrom et al. doubt the reliability of polarization techniques using Li metal electrodes for determining the parameters of liquid electrolytes.\\
Very-low-frequency impedance spectroscopy (VLF-IS) could isolate the convoluted information of the transference number and the thermodynamic factor from interfacial effects. While the current ratio method eliminates the solid-electrolyte interphase (SEI) and charge transfer resistance resonating at intermediate frequencies, it does not consider possible low-frequency diffusive effects through SEI or mossy Li.\cite{Single2019,Talian2019} These effects increase the measured steady-state potential and thus, can falsify the results for the two electrolyte parameters. Conducting VLF-IS down to very low frequencies may allow identifying and determining the diffusion resistance of the bulk electrolyte exclusively, revealing the desired undisturbed measurand.\\
Wohde et al. have already applied VLF-IS on symmetrical Li | electrolyte | Li cells.\cite{Wohde2016} Using dilute solution theory the group determined the transference number for three electrolytes for various electrode distances. However, the obtained transference numbers deviate strongly from the literature values of PFG- and PGSE-NMR measurements.\cite{Ueno2012, Porion2013, Niedzicki2011, Hofmann2015, Froemling2008} Also combining concentrated solution theory, the data of concentration cell measurements of Landesfeind et al.\cite{Landesfeind2019} and the VLF-IS measurements of Wohde et al.\cite{Wohde2016} does not yield matching results. To minimize convection and non-linear effects, Wohde et al. applied only small potentials leading to a tiny  concentration difference $\Delta c$ at the electrode sites, for instance $\Delta c < 0.77\,$mM for the LP30 electrolyte. These small concentration gradients could be one cause of the deviations.\\
Our paper focuses on determining the full set of electrolyte parameters for 0.5\,M LiPF$_6$ in EC:EMC (3:7, weight) at 20\,\textdegree C and 50\,\textdegree C, using a combination of concentration cell measurements, galvanostatic polarization experiments and electrochemical impedance spectroscopy (EIS). The subsequent section covers the corresponding theoretical framework, describing the relations between the involved measured observables and the desired parameters (see Section \ref{theory}). While section \ref{experimental} provides a detailed overview of the electrolyte preparation and the used experimental set-ups, we analyze the results in section \ref{results_and_discussion}. EIS measurements reveal the high-frequency resistance of the bulk electrolyte and enable the calculation of the corresponding conductivity. Concentration cell measurements yield convoluted data of the transference number and the thermodynamic factor. To deconvolute these quantities, the authors run galvanostatic polarization experiments with symmetrical metal Li | electrolyte + separator | Li metal cells. We discuss the deviation of the measured potential response from theoretical expectations and analyze the occurring effective diffusion processes. To spectrally resolve these processes we conduct VLF-IS with elevated current amplitudes leading to significantly high concentration gradients $\Delta c$.

\section{Theory and Experiments}
\subsection{Theoretical Background}
\label{theory}
In this work, we focus on determining the four electrolyte parameters of concentrated binary electrolytes containing LiPF$_6$ and organic solvents in a symmetric Li | electrolyte + separator | Li battery. The temporal evolution of the salt concentration $c=c_+=c_-$ of such a system (i.e. the stoichiometric coefficients of the salt are $\nu_+=\nu_-=1$) can be described by the continuity equation introduced by Latz et al.,\cite{Latz2011, Latz2015} with $c_i$ indicating the concentration of ion species $i$. Suppressing convection effects by filling the electrolyte in microporous separators and assuming the absence of chemical reactions in the bulk electrolyte, the continuity equation for the superficial phase reads
\begin{equation}
\label{continuity_equation}
\varepsilon\frac{\partial c}{\partial t} = -\nabla N = \nabla \left(\varepsilon^\beta D_\pm\left(c\right)\nabla c\right)-\nabla\left(\frac{t_+\left(c\right)}{z_+ F}\bar{i}\right).
\end{equation}
$\varepsilon$ and $\beta$ represent the porosity of the separator and the corresponding Bruggemann-coefficient, $D_\pm\left(c\right)$ is the salt diffusion coefficient, $t_+\left(c\right)$ and $z_+$ are the transference number and the charge number of the cation, and $F$ is the Faraday constant. $\bar{i}$ denotes the current density for the superficial phase
\begin{equation}
\label{current_density}\begin{aligned}
\bar{i} &= -\varepsilon^\beta \kappa\left(c\right)\nabla\varphi + \frac{RT}{z_+F}\varepsilon^\beta \kappa\left(c\right)\\
&\quad\cdot\left[1+\frac{\partial\ln f_\pm\left(c\right)}{\partial\ln c}\right]\left(1-t_+\left(c\right)\right)\frac{\nabla c}{c}
\end{aligned}
\end{equation}
where $\varphi$ indicates the electrochemical potential of the solution phase with respect to a lithium metal reference electrode, $R$ the ideal gas constant, $\kappa\left(c\right)$ the conductivity, and $1+\frac{\partial\ln f_\pm\left(c\right)}{\partial\ln c}=TDF$ the thermodynamic factor with the salt activity coefficient $f_\pm\left(c\right)$ of the electrolyte. $f_\pm\left(c\right)$ describes the deviation of the electrolyte system from ideal behavior and relates the chemical potential with the concentration gradient. The charge density is conserved \cite{Newman2004}
\begin{equation}
\label{charge_conservation}
\nabla\bar{i} = 0.
\end{equation}
Using a large active electrode surface area to electrode distance ratio allows reducing the differential equations to one dimension since the potential and concentration gradients are aligned perpendicular with respect to the parallel aligned electrodes.\cite{Ehrl2017a} In order to determine the electrolyte parameters, often experiments with only small concentration gradients compared to the bulk concentration $c_0$ are conducted. Doing so, the concentration-dependent parameters can be approximated in zeroth order to be constant for the small concentration fluctuations. Together with assuming a constant porosity $\varepsilon$ and Bruggemann-coefficient $\beta$ and inserting Eq. \ref{charge_conservation}, Eq. \ref{continuity_equation} then simplifies into
\begin{equation}
\label{simp_mass_conservation}
\varepsilon\frac{\partial c}{\partial t} =  \varepsilon^\beta D_\pm\left(c_0\right)\Delta c.
\end{equation}
The form of Eqs. \ref{current_density} and \ref{simp_mass_conservation} allow directly determining $\kappa\left(c_0\right)$ and  $D_\pm\left(c_0\right)$ using high-frequency impedance and polarization measurements, respectively (see Sections \ref{sec_VLF-IS} and \ref{theo_diff}). However, $t_+\left(c_0\right)$ and $TDF\left(c_0\right)$ are convoluted. To deconvolute, we use concentration cells, galvanostatic polarization experiments and VLF-IS yielding information in the form of factor $a\left(c_0\right)$ and $b\left(c_0\right)$ \cite{Landesfeind2019}
\begin{equation}
\label{factor_a}
a\left(c_0\right) = TDF\left(c_0\right)(1-t_+\left(c_0\right)),
\end{equation}
\begin{equation}
\label{factor_b}
b\left(c_0\right) = TDF\left(c_0\right)(1-t_+\left(c_0\right))^2.
\end{equation}
Combining both factors isolates $t_+\left(c_0\right)$ and $TDF\left(c_0\right)$ with
\begin{equation}
\label{decon_t}
t_+\left(c_0\right)=1-\frac{b\left(c_0\right)}{a\left(c_0\right)},
\end{equation}
\begin{equation}
\label{decon_TDF}
TDF\left(c_0\right) = \frac{a\left(c_0\right)^2}{b\left(c_0\right)}.
\end{equation}

\subsubsection{Concentration Cell}
\label{concentration_cell}
Concentration cells consist of two half-cells containing electrolytes with slightly different concentrations $c_0\pm\delta c$. In each solution, a Li metal electrode is immersed. The two half-cells are connected with a salt bridge to enable ion transfer and thus, to allow the measurement of the potential between the Li electrodes. However, the salt bridge is designed to allow only negligibly small current densities such that the current density approximates to $\bar{i} = 0$. With this, solving Eq. \ref{current_density} for $\frac{\partial\varphi}{\partial x}$ , integrating over $\mathrm dx$ and using the assumption of constant electrolyte parameters for small concentration gradients leads to an expression for the measured potential $U_\mathrm{conc}$ between the Li electrodes.\cite{Ehrl2017} Isolating the convoluted information of $t_+\left(c_0\right)$ and $TDF\left(c_0\right)$ yields
\begin{equation}
\label{conc_cell_equ}
a\left(c_0\right) = \frac{z_+F}{RT}\frac{U_\mathrm{conc}}{\ln\frac{c_0\ + \delta c}{c_0 - \delta c}}.
\end{equation}

\subsubsection{Galvanostatic Polarization}
\label{theo_diff}
\begin{figure*}[bp]%
  \centering
  \begin{subfigure}[t]{0.3\textwidth}
        \caption{} \label{theory_D_pot}
	\includegraphics[height=0.80\linewidth]{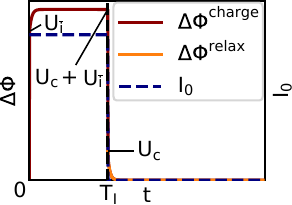}
  \end{subfigure}
  \qquad
  \begin{subfigure}[t]{0.3\textwidth}
        \caption{} \label{theory_D_sqrt}
	\includegraphics[height=0.80\linewidth]{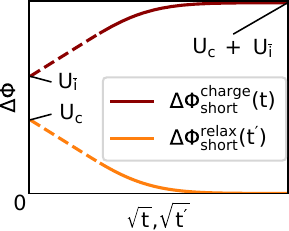}
  \end{subfigure}
  \qquad
  \begin{subfigure}[t]{0.3\textwidth}
  	\caption{} \label{theory_D_log}
	\includegraphics[height=0.80\linewidth]{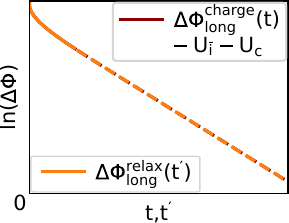}
  \end{subfigure}
  \caption{\textbf{a)} Theoretical potential response of a galvanostatic polarization experiment. The potentials $U_{\bar{i}}$, $U_{\bar{i}}+U_\mathrm{c}$ and $U_\mathrm{c}$ can be read off at the beginning, at the end and directly after the current pulse. \textbf{b)} Potential response during the charging and relaxation phase over $\sqrt{t}$ and $\sqrt{t^\prime}$ respectively. For short times, the potential exhibits a linear slope (dashed lines). \textbf{c)} Potential response $\ln\left( \Delta\Phi_\mathrm{long}^\mathrm{charge}-U_{\bar{i}} - U_\mathrm{c}\right)$ and $\ln\left(\Delta\Phi_\mathrm{long}^\mathrm{relax}\right)$. Both terms show the same response. For long times, the logarithmic terms exhibit again a linear slope (dashed lines).}
  \label{theory_D}
\end{figure*}
Galvanostatic pulse polarization experiments in symmetric Li | electrolyte + separator | Li cells allow the determination of the diffusion coefficient $ D_\pm$ and factor $b\left(c_0\right)$ (see Figure \ref{theory_D}). In these experiments a constant current density $\bar{i} = \frac{I_0}{A}$ with the amplitude $I_0$ and the active electrode area $A$ induces a gradient to the initially constant concentration profile of the cell. This establishes a potential $\Delta\Phi$ between the Li electrodes. For long enough polarization durations, the concentration gradient reaches a linear steady-state. After the current pulse, the concentration potential relaxes back to the ground state. The authors measure $\Delta\Phi$ during both the charging and the subsequent relaxation process. The derivation of the following equations is based on Refs.~\mbox{[\!\!\!\citenum{Newman1973, Hiller2013, Ehrl2017a, Ehrl, Ehrl2017}]}.\\
To induce the polarizing current we apply an overpotential $\eta$ on the Li electrodes. For small $\eta$, a linearized Butler-Volmer rate expression describes the reaction rate on the electrode surface, connecting the current density and the overpotential $\eta_\mathrm{lin}$ with the charge transfer resistance  $R_\mathrm{CT}$. For metallic Li electrodes, the linearized overpotential reads\cite{Single2019}
\begin{equation}
\label{overpot}
\eta_\mathrm{lin} = -R_\mathrm{CT}\bar{i} = \Delta\Phi - \Delta\varphi_\mathrm{bulk}
\end{equation}
with $\Delta\varphi_\mathrm{bulk}$ indicating the electrochemical potential of the bulk electrolyte. We express $\Delta\varphi_\mathrm{bulk}$ by solving Eq. \ref{current_density} for $\Delta\varphi$ and integrating over the full length $L$ of the cell. For small concentration differences $\Delta c$ with the approximation $\ln\frac{c +\delta c}{c -\delta c} \approx \frac{2\delta c}{c}= \frac{\Delta c}{c}$,\cite{Bruce1987} this yields 
\begin{equation}
\label{phi_bulk}\begin{aligned}
\Delta\varphi_\mathrm{bulk} &= -\frac{\bar{i}L}{\varepsilon^\beta \kappa\left(c_0\right)} + \frac{RT}{z_+F}\\
&\quad\cdot TDF\left(c_0\right)\left(1-t_+\left(c_0\right)\right)\frac{\Delta c}{c_0}.
\end{aligned}
\end{equation}
In order to calculate $\Delta c$, we solve the diffusion equation Eq. \ref{simp_mass_conservation}, yielding the concentration profile $c\left(x,t\right)$. For this, we consider the boundary conditions at the electrodes. During the charging process, the induced current density leads to a net Li-ion flux at the electrodes $\bar{i}=z_+ FN$ as the anions are blocked. Together with Eq. \ref{continuity_equation}, this yields \cite{Newman2004}
\begin{equation}
\label{boundary}
\frac{\partial c}{\partial x} = -\frac{1-t_+\left(c_0\right)}{z_+F \varepsilon^\beta D_\pm\left(c_0\right)}\bar{i}.
\end{equation}
Solving the diffusion equation Eq. \ref{simp_mass_conservation} for an initially constant concentration profile $c\left(x,0\right) = c_0$ approximates the temporal evolution of $\Delta c$ to a square-root function for small charging times $t$. For long charging times, the concentration difference converges exponentially to its final value. The respective measurable potentials thus read 
\begin{equation}
\Delta\Phi_\mathrm{short}^\mathrm{charge}= U_{\bar{i}} + U_\mathrm{c}\sqrt{\frac{16D_\pm\left(c_0\right) t}{\varepsilon^{1-\beta}\pi L^2}}
\end{equation}
and
\begin{equation}
\Delta\Phi_\mathrm{long}^\mathrm{charge}=U_{\bar{i}} + U_\mathrm{c}\left(1-\frac{8}{\pi^2}\mathrm{e}^{-\frac{\pi^2}{\varepsilon^{1-\beta} L^2}D_\pm\left(c_0\right) t}\right).
\end{equation}
Here, $U_{\bar{i}}$ summarizes the potential contributions depending on the current density $\bar{i}$. As the applied charging pulses are galvanostatic $U_{\bar{i}}$ is constant over time. Solely the temporal evolution of $\Delta c$ dictates the shape of $\Delta\Phi$. For $t\rightarrow\infty$, $\Delta c$ establishes a linear profile and the concentration potential difference converges to the steady-state $U_\mathrm{c}$.
\begin{equation}\begin{aligned}
\label{U_c}
 U_\mathrm{c} &= -\frac{RT}{z_+^2F^2}b\left(c_0\right)\frac{\bar{i}}{c}\frac{L}{\varepsilon^\beta D_\pm\left(c_0\right)}\end{aligned}
\end{equation}
As evident in Eq. \ref{U_c}, $U_\mathrm{c}$ comprises factor $b\left(c_0\right)$ (see Eq. \ref{factor_b}). Therefore, measuring $U_\mathrm{c}$ allows specifying $b\left(c_0\right)$ for known $D_\pm\left(c_0\right)$. The corresponding resistance $R_\mathrm{c} = \frac{U_\mathrm{c}}{I_0}$ can also be measured in VLF-IS measurements (see Eq. \ref{equ_RD}). Combining factor $b\left(c_0\right)$ with factor $a\left(c_0\right)$ from the concentration cell measurements allows isolating $t_+\left(c_0\right)$ and $TDF\left(c_0\right)$.\\
After the polarization pulse, the current is switched off ($\bar{i}(t>T_\mathrm{I}) = 0$) and the built up concentration potential relaxes to the ground state. In order to calculate the time evolution of the corresponding potential $\Delta\Phi$, we again solve the diffusion equation Eq. \ref{simp_mass_conservation}. Since there is no net Li on flux at the electrodes the boundary conditions read $\frac{\partial c}{\partial x} = 0$. Assuming a linear concentration profile at the current interruption time $T_\mathrm{I}$ leads to similar solutions for the relaxing process as for the charging process. Due to the absence of an applied current density, $U_{\bar{i}}$ vanishes. For short and long times $t^\prime=t-T_\mathrm{I}$, the approximated potentials read \cite{Ehrl2017a, Hiller2013}
\begin{equation}
\Delta\Phi_\mathrm{short}^\mathrm{relax}=  U_\mathrm{c}\left(1-\sqrt{\frac{16D_\pm\left(c_0\right) t^\prime}{\varepsilon^{1-\beta}\pi L^2}}\right)
\end{equation}
and
\begin{equation}
\Delta\Phi_\mathrm{long}^\mathrm{relax}=U_\mathrm{c}\frac{8}{\pi^2}\mathrm{e}^{-\frac{\pi^2}{\varepsilon^{1-\beta} L^2}D_\pm\left(c_0\right) t^\prime}.
\end{equation}
Plotting the short-term potential behavior of the charge and discharge process over $\sqrt{t}$ and $\sqrt{t^\prime}$ yields a linear slope $m_\mathrm{sqrt}$ (see Figure \ref{theory_D_sqrt}). Therefore, measuring this slope in the polarization experiment allows determining $D_\pm^\mathrm{sqrt}\left(c_0\right)$ with
\begin{equation}
\label{D_sqrt}
D_\pm^\mathrm{sqrt}\left(c_0\right) = \frac{\pi L^2}{16}\left(\frac{m_\mathrm{sqrt}}{U_\mathrm{c}}\right)^2\varepsilon^{1-\beta}.
\end{equation}
For the long-term behavior, we plot $\ln\left( \Delta\Phi_\mathrm{long}^\mathrm{charge}-U_{\bar{i}} - U_\mathrm{c}\right)$ and $\ln\left(\Delta\Phi_\mathrm{long}^\mathrm{relax}\right)$ over $t$ and $t^\prime$ respectively (see Figure \ref{theory_D_log}). This again leads to a linear slope $m_\mathrm{ln}$, relating to $D_\pm^\mathrm{ln}\left(c_0\right)$ with
\begin{equation}
\label{D_ln}
D_\pm^\mathrm{ln}\left(c_0\right) = \frac{L^2}{\pi^2}m_\mathrm{ln}\varepsilon^{1-\beta}.
\end{equation}
The potentials $U_{\bar{i}}$ and $U_\mathrm{c}$ can be read off at the beginning or immediately after the charging process when either $U_\mathrm{c}=0$ or $U_{\bar{i}}=0$ (see Figures \ref{theory_D_pot} and \ref{theory_D_sqrt}). However, at the beginning of the charging and relaxation process, the slope of $\Delta\Phi$ is steep. Therefore, the corresponding data points are prone to measurement errors. Using EIS to measure the bulk and interfacial resistance $R_\mathrm{el}$ and $R_\mathrm{int}$ before or during the polarization process provides a more precise way to determine $U_{\bar{i}} = I_0\left(R_\mathrm{el}+R_\mathrm{int}\right)$. Thereby, $R_\mathrm{int}$ comprises the charge transfer resistance $R_\mathrm{CT}$ and the resistance of the SEI $R_\mathrm{SEI}$. Measuring the potential during the steady-state at the end of the charging process and subtracting $U_{\bar{i}}$ yields $U_\mathrm{c}$.\\

\subsubsection{Very-Low-Frequency Impedance Spectroscopy}
\label{sec_VLF-IS}
For the theoretical description of EIS, the authors refer to the paper of Single et al.,\cite{Single2019} deriving the impedance equations with the transport theory of Schammer et al.\cite{Schammer2021} Schammer et al. derive a holistic continuum theory covering the characteristic phenomena of various multi-component solutions near electrified interfaces and in the bulk.\cite{Schammer2021} Compared to the model from Latz et al.,\cite{Latz2011, Latz2015} this theory introduces several correction factors making the model even more precise. Single et al. use this theory to develop a detailed, physics-based model for impedance spectroscopy.\cite{Single2019} The following section gives a rough overview of the derived analytical equations. For LiPF$_6$ in EC:EMC (3:7, weight) the correction factors are small. Therefore, we neglect these factors and continue with an analog derivation using the model from Latz et al. presented above.\\
Galvanostatic electrochemical impedance spectroscopy measures the resulting potential difference of cathode and anode $\Delta \Phi$ of a system induced by an oscillating current density $\bar{i} = \frac{I_0}{A}\mathrm{e}^{i\omega t}$ with the angular frequency $\omega$. For our symmetrical cell, the impedance reads
\begin{equation}
\label{impedance}
Z\left(\omega\right) = \frac{\Delta \Phi}{I_0\mathrm{e}^{i\omega t}}.
\end{equation}
$\Delta\Phi$ depends on the overpotential $\eta_\mathrm{lin}$ and the electrochemical potential $\Delta\varphi_\mathrm{bulk}$ (see Eqs. \ref{overpot} and \ref{phi_bulk}). To find an expression for $\Delta c$ we solve Eq. \ref{simp_mass_conservation} with the antisymmetric ansatz $c\left(x, t\right) = C\mathrm{e}^{i\omega t}\sin kx$ since symmetric solutions in $x$ do not contribute to the impedance. This yields the wave vector $k$
\begin{equation}
k = \left(1-i\right)\sqrt{\frac{\varepsilon^{1-\beta}\omega}{2D_\pm\left(c_0\right)}}.
\end{equation}
The boundary conditions (see Eq. \ref{boundary}) at the electrodes $x=\pm\frac{L}{2}$ define the amplitude $C$. Inserting the expression into the ansatz results in 
\begin{equation}
c\left(x, t\right) = \frac{I_0\mathrm{e}^{i\omega t}}{z_+FA}\frac{1-t_+\left(c\right)}{\varepsilon^\beta D_\pm\left(c\right)}\frac{\sin kx}{k\cos \frac{kL}{2}}.
\end{equation}
Using the concentration difference at the electrodes $\Delta c = c\left(\frac{L}{2},  t\right) - c\left(-\frac{L}{2}, t\right)$ allows determining $\Delta\Phi$ and calculating the impedance with Eq. \ref{impedance}. This yields $Z\left(\omega\right) = Z_\mathrm{el}+Z_\mathrm{CT}+Z_\mathrm{D}$ including three terms accounting for the impedance of the bulk electrolyte in the separator $Z_\mathrm{el}$, the charge transfer impedance $Z_\mathrm{CT}$ and the corresponding diffusion impedance $Z_\mathrm{D}$.\\
In reality, the impedance $Z_\mathrm{CT}$ and $Z_\mathrm{el}$ show complex contributions due to capacitive behavior. However, this can only be explained in a non-local electro-neutral framework (see Single et al.\cite{Single2019}). In our simplified local electro-neutral framework $Z_\mathrm{CT}$ and $Z_\mathrm{el}$ are reduced to their real contributions $R_\mathrm{CT}$ and 
\begin{equation}
\label{Rel_equ}
R_\mathrm{el} = \frac{L}{\varepsilon^\beta \kappa\left(c_0\right) A}.
\end{equation} 
Conductivity cells allow determining $R_\mathrm{el}$ of the sole electrolyte $\left(\varepsilon^\beta=1\right)$ at high frequencies $\omega$, as the cells prevent any Li-flux through the blocking electrodes. The quotient $\frac{L}{A}$ is denoted as the cell constant $k$. For a non-trivial geometry, $k$ has to be determined with an electrolyte of known conductivity. This enables the evaluation of $\kappa\left(c_0\right)$.\\
Measuring $R_\mathrm{el}$ in Li | electrolyte + separator | Li cells and combining it with the known conductivity of the electrolyte reveals the MacMullin number of the separator $N_\mathrm{M}=\varepsilon^\beta$. This enables calculating the Bruggemann-coefficient $\beta$, as the porosity $\varepsilon$ is often stated by the manufacturer or can be measured separately. Both of these quantities are needed for the evaluation of the transport numbers.\\
The amplitude $R_\mathrm{D}$ of the diffusion impedance $Z_\mathrm{D}$ contains information about the thermodynamic factor and the transference number in the form of factor $b\left(c_0\right)$ (see Eq. \ref{factor_b}). 
\begin{equation}
Z_\mathrm{D} = R_\mathrm{D}\frac{\tan k\frac{L}{2}}{k\frac{L}{2}}
\end{equation}
\begin{equation}
\label{equ_RD}
R_\mathrm{D} = \frac{z_+^2RT}{F^2c}\frac{L}{A\varepsilon^\beta D_\pm\left(c_0\right)}b\left(c_0\right)
\end{equation}
Together with factor $a\left(c_0\right)$ measured by concentration cells, this allows deconvoluting $TDF\left(c_0\right)$ and $t_+\left(c_0\right)$. Although we do not model it here, diffusion through the SEI and possibly grown mossy Li on the pristine Li surface may have similar resonances as the diffusion through the separator.\cite{Single2019, Talian2019} Thus, it is important to be able to distinguish and identify the corresponding features in the impedance measurement.\\
Therefore, we numerically calculate the resonance frequency $f_\mathrm{res}$ of $Z_\mathrm{D}$ with a modified Newton-Raphson iteration method.\cite{CruzManzo2020} The resulting $f_\mathrm{res}$ depends on the diffusion coefficient $D_\pm\left(c_0\right)$ and the length $L$ of the separator. This allows determining $D_\pm\left(c_0\right)$ by measuring $f_\mathrm{res}$
\begin{equation}
f_\mathrm{res} = \frac{1.2703D_\pm\left(c_0\right)}{\pi\varepsilon^{1-\beta} \left(\frac{L}{2}\right)^2}.
\end{equation}
However, fractional structures and altered surface geometries may influence the impedance data.\cite{Keiser1976} In the literature, the data is therefore often fitted with a slightly modified expression for the Warburg short element. Similar to constant phase elements, the modification fits diffusion impedance data better, which end in a depressed semi-circle at low frequencies.
\begin{equation}
\label{Warburg_modified}
Z\left(\omega\right) = R_\mathrm{D}\frac{\tanh\left(is\right)^\alpha}{\left(is\right)^\alpha}
\end{equation}
The expression only coincides for $\alpha = 0.5$ with our model. For varying $\alpha$, the resonance frequency $f_\mathrm{res}$ shifts. A direct determination of $D_\pm\left(c_0\right)$ may therefore be erroneous.

\subsection{Experimental Methods}
\label{experimental}
The cell assembly and electrolyte preparation were done in argon-filled glove boxes (Jacomex GPT4FF, <1 ppm H$_2$O, <3 ppm O$_2$ and GS Glovebox Systemtechnik GmbH MEGA E-Line, <1 ppm H$_2$O, <1 ppm O$_2$). All cell parts and electrolyte mixing tools were dried overnight in a heat-able airlock attached to the glove box at different temperatures between 60\,\textdegree C and 120\,\textdegree C.\\
For preparing the electrolytes, the authors dissolved ethylene carbonate (EC, Alfa Aesar, anhydrous 99\%) in  ethyl-methylene carbonate (EMC, Solvionic, battery grade) in a 3:7 weight ratio, using a Mettler Toledo balance (AB135-S/FACT DualRange Analytical Balance). After a few hours of stirring, we mixed the solvent with different amounts of weighted lithium hexafluorophosphate (LiPF$_6$, Solvionic, 99.99\% battery grade) in volumetric flasks resulting in the salt concentrations 0.25, 0.5 and 0.75\,M. The blends were subsequently stirred for another several hours. The electrodes used for the concentration cell and polarization experiments consisted of metallic lithium (Alfa Aesar, 750\,\textmu m thickness, 99.9\% purity and Sigma-Aldrich, 380\,\textmu m thickness, 99.9\% purity) with a clean scraped surface.
\begin{figure*}[bp]%
  \centering
  \begin{subfigure}[t]{0.45\textwidth}
  	\caption{} \label{conc_pic_set-up}
	\includegraphics[width=\linewidth]{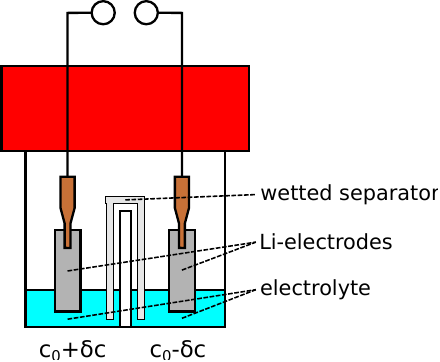}
  \end{subfigure}
  \qquad
  \begin{subfigure}[t]{0.45\textwidth}
  	\caption{} \label{gal_pol_set-up}
	\includegraphics[width=\linewidth]{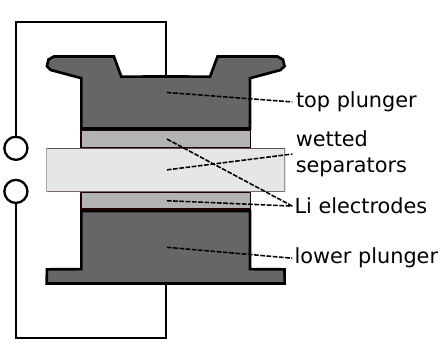}
  \end{subfigure}
  \caption{\textbf{a)} Concentration cell, filled with two different concentrations. The soaked separator allows a small current to flow and thus, the measurement of the potential difference between the two immersed Li electrodes. \textbf{b)} Cell stack for galvanostatic polarization experiments and very-low-frequency impedance spectroscopy (VLF-IS) measurements.} \label{conc_cell_label}
\end{figure*}\\
All experiments were conducted in a climate chamber (Vötsch Industrietechnik GmbH, LabEvent L T/64/40/3), providing a stable temperature ($\pm1$\,\textdegree C). Zennium Pro and Zahner IM6 potentiostats (Zahner-Elektrik GmbH) recorded the experimental data. For testing the obtained EIS spectra for linearity with Kramers-Kronig and fitting the  data with equivalent circuits we used the RelaxIS software (rhd instruments).

\subsubsection{Conductivity Cell}
\label{exp_sec_cond_cell}
Using the airtight TSC 1600 closed cell (rhd instruments) allows determining the conductivity of the 0.5\,M EC:EMC (3:7, weight) electrolyte. The cell consists of a platinum crucible and platinum electrodes. We filled the crucible with roughly 1\,ml of electrolyte, put the cell into the climate chamber and conducted every 5\,min potentiostatic EIS measurements from 8\,MHz to 5\,kHz with an amplitude of 30\,mV for 2.5\,h. This allows determining the point in time when the electrolyte within the cell has reached thermal equilibrium, as the determined high-frequency resistance becomes constant over time. Averaging several subsequent resistance values of two identical cells yields the measuring result. In order to translate the measured impedance into conductivity values we specified the cell constant $k$, comprising the cell's geometry. For this, we conducted the same experiments with a reference electrolyte (Conductivity Standard ROTI®Calipure 12880\,\textmu S/cm (25\,\textdegree C)) with known conductivity (see SI Section 1). For each calculated conductivity data point, an estimated error of 3\% accounts for the uncertainty of the cell constant, the concentration, the temperature and the fit.

\subsubsection{Concentration Cell}
\label{exp_sec_conc_cell}
We used custom concentration cells, fabricated by the glass-blowing of the University of Stuttgart (see Figure \ref{conc_pic_set-up}) to determine factor $a\left(c_0\right)$. The cells consist of a GL 45 threaded glass pipe with an attached bottom plate. An additional inserted glass plate separates two halves of the pipe, which contain several ml of electrolyte with different concentrations $c=c_0\pm\delta c$. Connecting both half-cells with a slim, over 5\,cm long piece of porous polyethylene (Nitto SUNMAP LC, 500\,\textmu m thickness, 0.3 porosity) enabled a small ionic current. Thereby, each half of the separator was soaked with the corresponding electrolyte concentration. Together with its Bruggemann-coefficient $\beta = 2.30$,\cite{Landesfeind2019} the long dimension of the separator ensured the concentrations in the half-cells to hardly vary due to diffusion during the measurement time of several hours. In fact, modeling the diffusion (see Section \ref{model}) suggests a change of concentration difference of less than 1\textperthousand\ within the first 13 days at 50\,\textdegree C and even less at 20\,\textdegree C. In order to measure the potential between the half-cells, we connected brushed Li metal electrodes to crocodile clamps and immersed the electrodes into the electrolyte. Cleaning the surface of the Li electrodes seems to be crucial since experiments without brushing lead to unstable measured potentials and fast potential drops, even if the electrodes appear clean and shiny. A threaded cap with a septum allowed to air-tightly close the cell and connect the Li electrodes with the measuring device.\\
We measured the open-circuit potential for the base concentration of $c_0 = 0.5\,$M for several hours at 20\,\textdegree C and 50\,\textdegree C and averaged the results of two identical cells. The long measurement duration ensured that the cells had reached the desired temperature. Choosing $\delta c = 0.25$\,M led to concentrations of 0.25\,M and 0.75\,M in the corresponding half-cells. For the error calculation, we estimated an error of 3\% for $\delta c$ and an error of $1\,\textdegree$C for $T$.

\subsubsection{Galvanostatic Polarization}
\label{exp_diffusion}
We conducted the galvanostatic pulse polarization experiments with symmetric Li metal | electrolyte + separator | Li metal systems using ECC-PAT-Core cells from El-Cell GmbH. The separator of each cell consisted of 20 punched polypropylene Celgard 2500 layers (CG, 25\,\textmu m thickness, 0.55 porosity) with a diameter of 21.6\,mm and a total thickness of 500\,\textmu m. Immersing the separators in the 0.5\,M LiPF$_6$ in EC:EMC (3:7, weight) electrolyte overnight ensured sufficient wetting. The total height of the used cell stack of Li electrodes and separators would exceed the intended limits of the cell. Therefore, we tapped the lower and upper Li electrodes (18\,mm diameter, 750\,\textmu m and 380\,\textmu m thickness) to the corresponding lower (size 800) and upper plunger to thin them out. We stacked the wetted separators in the insulation sleeve and inserted the plungers with the electrodes. Subsequently putting the stack into the cell housing and closing the cell air-tightly with the seal ring finished the cell assembly. Figure \ref{gal_pol_set-up} shows a sketch of the used cell stack.
\begin{table}[bt]
	\begin{center}
	\caption{Measuring schedule for the galvanostatic polarization experiments and the VLF-IS measurements. Before the actual measurements (Meas.) the cells run through a conditioning process (Cond.). Between each galvanostatic polarization experiment, we conduct EIS measurements, which are not listed here.} \label{ac_schedule}
		\begin{tabular}{ccc}	
\toprule		
 & Polarization amplitudes & VLF-IS amplitudes\\
\midrule
Cond. & 2x 10\,\textmu A (VLF-IS) & \multirow{2}{*}{3x 10$\,\textmu$A}\\%
20\,\textdegree C & 2x 40\,\textmu A (VLF-IS) & \\
& & \\
	& $\pm 10$\,\textmu A & \multirow{2}{*}{2x 10\,\textmu A}\\
\multirow{2}{*}{Meas.} & $\pm 20$\,\textmu A & \multirow{2}{*}{4x 20\,\textmu A}\\
\multirow{2}{*}{20\,\textdegree C} & $\pm 40$\,\textmu A & \multirow{2}{*}{2x 40\,\textmu A}\\
	& $\pm 60$\,\textmu A & \multirow{2}{*}{2x 60\,\textmu A}\\
	& $\pm 10$\,\textmu A & \\
 & & \\
 \multirow{2}{*}{Cond.} & 1x 20\,\textmu A (VLF-IS) & 1x 20\,\textmu A\\
\multirow{2}{*}{50\,\textdegree C} & 2x 50\,\textmu A (VLF-IS) & 2x 50\,\textmu A\\
	& 3x 100\,\textmu A (VLF-IS) & 3x 100\,\textmu A\\
  & & \\
  	& $\pm 5$\,\textmu A & 2x 5\,\textmu A\\
	& $\pm 10$\,\textmu A & 2x 10\,\textmu A\\
	& $\pm 25$\,\textmu A & 2x 25\,\textmu A\\
	& $+10$\,\textmu A & 10\,\textmu A\\
	& $\pm 50$\,\textmu A & 2x 50\,\textmu A\\
	& $+10$\,\textmu A & 10\,\textmu A\\
	& $\pm 75$\,\textmu A & 2x 75\,\textmu A\\
	& $+10$\,\textmu A & 10\,\textmu A\\
	& $\pm 100$\,\textmu A & 2x 100\,\textmu A\\
Meas. & $+10$\,\textmu A & 10\,\textmu A\\
50\,\textdegree C & $\pm 150$\,\textmu A & 2x 150\,\textmu A\\
	& $+10$\,\textmu A & 10\,\textmu A\\
	& $\pm 200$\,\textmu A & 2x 200\,\textmu A\\
	& $+10$\,\textmu A & 10\,\textmu A\\
	& $\pm 300$\,\textmu A & 2x 300\,\textmu A\\
	& $+10$\,\textmu A & 10\,\textmu A\\
	& $\pm 400$\,\textmu A & 2x 400\,\textmu A\\
	& $+10$\,\textmu A & 10\,\textmu A\\
	& $\pm 500$\,\textmu A & 2x 500\,\textmu A\\
	& $+10$\,\textmu A & 10\,\textmu A\\
\bottomrule	
	\end{tabular}
	\end{center}

\end{table}\\
Before the first measurements, the cells ran through a conditioning process at a temperature of 20\,\textdegree C to stabilize the SEI. The process consisted of three VLF-IS from 4\,MHz to 400\,\textmu Hz with three measuring periods, seven steps per decade, and current amplitudes of $I_0=10$\,\textmu A. After the conditioning we galvanostatically polarized the cells for 1\,h and tracked the subsequent potential relaxation for 9\,h, using current amplitudes from 10\,\textmu A - 60\,\textmu A with alternating signs. The long charging time ensured establishing the steady-state and thus a linear concentration gradient. After each experiment, we conducted potentiostatic EIS from 4\,MHz to 1\,Hz with an amplitude of 5\,mV to track any changes to the interfacial resistance. Additionally, the bulk resistance at high frequencies together with the previously measured conductivity of the sole electrolyte reveals the Bruggemann-coefficient of the separator.\\
For the subsequent measurements at 50\,\textdegree C, the cells ran through the very same conditioning and polarization procedure but with elevated current amplitudes. The amplitudes ranged from 20\,\textmu A to 100\,\textmu A for the VLF-IS conditioning and from 5\,\textmu A to 500\,\textmu A for the galvanostatic polarization experiments. These experiments required higher potentials and had an overall duration of over 220\,h. Therefore, we additionally conducted reference polarization measurements with $I_0 = 10$\,\textmu A between the measurements with elevated current amplitudes to track any changes in the system. A detailed measuring schedule is given in Table \ref{ac_schedule}.\\
To calculate $D_\pm\left(c_0\right)$ we evaluated the short- and long-term behavior of the potential and averaged the results of three identical cells.

\subsubsection{Very-Low-Frequency Impedance Spectroscopy}
For the VLF-IS we used the same cell structure, conditioning and measuring procedure as in the galvanostatic pulse experiments (see Section \ref{exp_diffusion}). However, the total separator thickness has to be chosen with care. Stacking more separators in the insulation sleeve increases the diffusion resistance $R_\mathrm{D}$ allowing a more accurate measurement of this quantity. Simultaneously, a thicker separator size shifts the system's resonance to lower frequencies extending the time of the VLF-IS measurement. Considering the transport parameters of the electrolyte at 20\,\textdegree C and 50\,\textdegree C, the porosity and Bruggemann-coefficient reported in the literature\cite{Landesfeind2019, Landesfeind2016} the authors decided to stack 20 CG separators. This  results in a total separator thickness of 500\,\textmu m as in the galvanostatic pulse experiments. The thickness maximizes the calculated diffusion resistance to a value of 13.21\,$\Omega$ and 5.41\,$\Omega$ with a corresponding resonance of 0.92\,mHz and 1.63\,mHz. This allows measurements with less than 10\,h duration.\\
The VLF-IS experiment started with a similar conditioning procedure as in the polarization measurements at 20\,\textdegree C, followed by the actual VLF-IS measurements with current amplitudes $I_0 = 10$\,\textmu A - 60\,\textmu A. Subsequently, the cells ran through a second conditioning process at 50\,\textdegree C before the next set of VLF-IS measurements, applying current densities from 5\,\textmu A up to 500\,\textmu A. At this temperature, we conducted additional VLF-IS reference measurements with $I_0 = 10$\,\textmu A between the measurements with elevated amplitudes to track any changes to the cell. The measuring schedule is described in Table \ref{ac_schedule}.

\subsection{Simulation Set-up}
\label{model}
\begin{figure}[hbt]	
	\centering
	\includegraphics[width=1.0\linewidth]{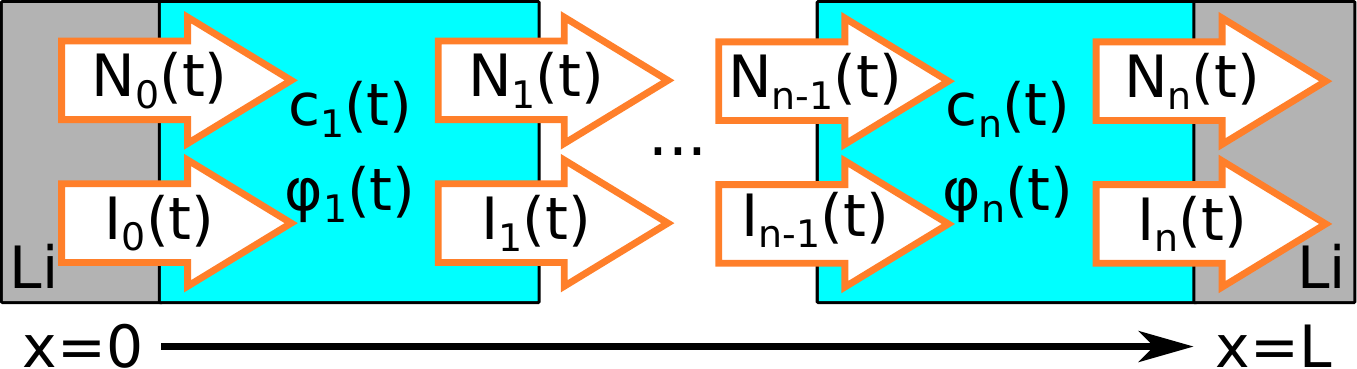} %
	\caption{Scheme of the 1-D model.}
	\label{model_fig}
\end{figure}

In order to predict the results of the experiments we use a 1-D model. The model calculates the spatially resolved temporal evolution of the concentration $c\left(x,t\right)$ and the electrochemical potential $\varphi\left(x,t\right)$ of the electrolyte within the separator with length $L$. For this, the separator soaked with electrolyte divides into $n=500$ segments of the same length. At $x=0$ and $x=L$ Li metal electrodes are located (see Figure \ref{model_fig}).\\
\\
Depending on the applied current density $\bar{i}$ the electrodes induce a net Li-ion flux $N=\frac{\bar{i}}{z_+\nu_+ F}$ into or out of the adjacent segments. Thereby, the standard Butler-Volmer equation connects the current density with the corresponding overpotential $\eta_i=\phi_i\left(t\right)-\varphi\left(x_i,t\right)$ ($i=\mathrm{anode}$, cathode)
\begin{equation}
\label{Butler-Volmer-equ.}
\bar{i}=\bar{i}_0\left[\exp\left(\frac{\alpha z_+F}{RT}\eta\right) - \exp\left(-\frac{\left(1-\alpha\right) z_+F}{RT}\eta\right)\right]
\end{equation}
with the exchange current density $\bar{i}_0$ and the symmetry factor $\alpha = 0.5$. The difference $\Delta\Phi = \phi_\mathrm{anode}-\phi_\mathrm{cathode}$ denotes the measurable potential difference between the Li electrodes.\\
Eqs. \ref{continuity_equation} - \ref{charge_conservation} describe the resulting current and Li-ion flux between the individual segments as well as the temporal evolution of the concentration $c_i\left(t\right)$ and the electrochemical potential $\varphi_i\left(t\right)$ ($i= 1, \ldots, n$) within each segment. We solve this differential-algebraic system of equations using the Matlab solver ode15s. The occurring electrolyte parameters are governed by the corresponding empirical approximations by Landesfeind et al.\cite{Landesfeind2019} for $c=0.5\,$M, $T=20$\,\textdegree C and $T=50$\,\textdegree C (see Table \ref{sim_param_table}). For modeling non-linear effects we use concentration-dependent parameters (see SI Section 4.1). Note, that Landesfeind et al. base the definition of the thermodynamic factor on the chemical potential of the Li-ions where we use the chemical potential of the neutral salt which is the sum of the chemical potential of anions and cations (weighted with the correct stoichiometric factors). Therefore, the $TDF$ of Landesfeind et al. has to be multiplied by 2 for being incorporated in our model.\\
A slight variation of the model can also predict the behavior of our concentration cells.

\begin{table}[htb]
	\begin{center}
	\caption{Constant electrolyte parameters for $c=0.5\,$M used in the 1-D model, taken from Landesfeind et al.\cite{Landesfeind2019}}\label{sim_param_table}
		\begin{tabular}{ccc}	
\toprule		
Parameter & 20\,\textdegree C & 50\,\textdegree C\\
\midrule
$\kappa\,$(mS/cm) 	&	 7.0	&	10.3\\
$D_\pm\cdot 10^{-10}\,$(m$^2$/s) 	&	 3.6	&	6.3\\
$t_+$ 	&	 0.3	&	0.5\\
$TDF$ 	&	 2.2\textsuperscript{[a]}	&	3.0\textsuperscript{[a]}\\
\bottomrule	
	\end{tabular}
	\end{center}
 \footnotesize{\textsf{[a] These values are multiplied by 2.}}
\end{table}

\section{Results and Discussion}
\label{results_and_discussion}

\subsection{Conductivity Cell}
\label{res_cond_cell}
\begin{figure}[!t]	
	\centering
	\includegraphics[width=0.9\linewidth]{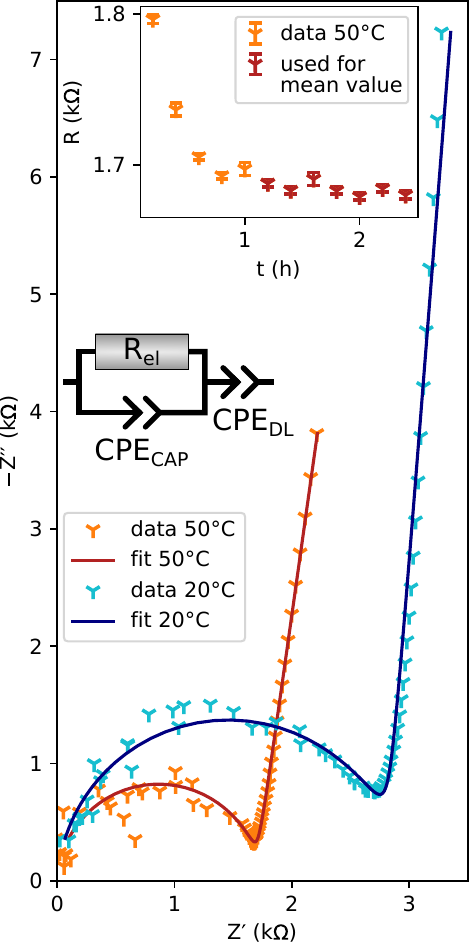} %
	\caption{Exemplary conductivity measurement. We fit the data determined by electrochemical impedance spectroscopy (EIS) measurements of the sole electrolyte with the equivalent circuit shown in one of the insets. The second inset shows the temporal evolution of $R_\mathrm{el}$ with the corresponding fit error. $R_\mathrm{el}$ fluctuates around a constant value after being tempered for roughly one hour in the climate chamber, indicating that the electrolyte reached its final temperature of 50\,\textdegree C.}
	\label{cond_pic}
\end{figure}

To determine the conductivity $\kappa\left(c_0\right)$ we measure the impedance of the sole 0.5\,M EC:EMC (3:7, weight) electrolyte within the conductivity cell from 8\,MHz to 5\,kHz located in a climate chamber. The data shows in the Nyquist plot a semi-circle at high frequencies, followed by a linear capacitive increase at lower frequencies.\\
We fit the data with an equivalent circuit consisting of a constant phase element (CPE) parallel to a resistance $\left(R_\mathrm{el}\right)$ and an additional CPE in series (see Figure \ref{cond_pic}). The parallel CPE and $R_\mathrm{el}$ represent the capacity of the conductivity cell and the resistance of the bulk electrolyte. The serial CPE describes the capacity induced by the double layer at the electrodes.\\
As the cell is reaching the desired temperature, $R_\mathrm{el}$ converges to its final value, indicating that the electrolyte has thermalized with the climate chamber (see Figure \ref{cond_pic} inset). Averaging several subsequent impedance measurements for two identical cells yields the final value of the bulk electrolyte resistance. The cell constant $k$ (see SI Section 1) maps the measured resistances to conductivities with 
\begin{equation}
\label{res_kappa_equ}
\kappa = \frac{k}{R_\mathrm{el}}.
\end{equation}
With this, the conductivity of 0.5\,M LiPF$_6$ in EC:EMC (3:7, weight) results in $\kappa\left(c_0\right) = 7.1\pm 0.2\,\frac{\mathrm{mS}}{\mathrm{cm}}$ at 20\,\textdegree C and $\kappa\left(c_0\right) = 9.6\pm 0.3\,\frac{\mathrm{mS}}{\mathrm{cm}}$ at 50\,\textdegree C, which resemble well the values found in the literature.\cite{Landesfeind2019} Further conductivity measurements with various concentrations and temperatures can be found in the supplementary (see SI Section 1).

\subsection{Concentration Cell}
\label{sec_conc_cell}
\begin{figure}[bt]	
	\centering
	\includegraphics[width=\linewidth]{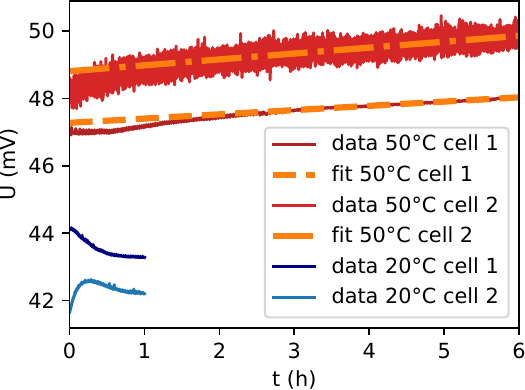} 
	\caption{Concentration cell measurement. At 20\,\textdegree C, the potential of both concentration cells takes constant values after roughly one hour, indicating thermal equilibrium with the climate chamber. For the measurements at 50\,\textdegree C, the potential difference linearly increases for several hours. To eliminate the slope we extrapolate the data to $t=0$. The noise difference of the cells originates from different voltage resolutions.}
	\label{conc_pic}
\end{figure}

To specify factor $a\left(c_0\right)$ (see Eq. \ref{factor_a}), we track the potential between the immersed Li electrodes of the two identical concentration cells over the heating/cooling process. Constant potentials should indicate thermal equilibrium of the 0.25\,M and 0.75\,M electrolytes within the half-cells with the climate chamber.\\
At 20\,\textdegree C, the equilibrium is reached after roughly one hour as the potentials of the cells have stabilized (see Figure \ref{conc_pic}). Subsequently averaging each potential for 5\, min and taking the mean of both results yields $U_\mathrm{conc}$.
\\
At 50\,\textdegree C, the potential for both cells slopes within the first 2-3\,h into a slowly linearly increasing behavior, which even remains until the end of the 6\,h measurement. The deviating noise levels between the two cells at 50\,\textdegree C originate from using two different measurement devices with different voltage resolutions for each cell.\\
The origin of this increasing potential is not quite clear. The balancing of the concentration difference between the two half-cells through the separator would lead to a decreasing potential at much larger time scales. Therefore, the authors suspect chemical reactions to alter the concentrations in the vicinity of the electrodes at different rates. Also, different evaporation rates of the electrolyte could lead to an increase in the concentration difference.\\
To eliminate the impact of the unknown effect, we linearly fit the potential and calculate the potential difference at $t = 0$ (see Figure \ref{conc_pic}). Averaging both results yields the mean potential difference $U_\mathrm{conc}$ for the two cells.\\
Inserting $U_\mathrm{conc}$ in Eq. \ref{conc_cell_equ} reveals the factor $a\left(c_0\right) = 1.54\pm 0.06$ at 20\,\textdegree C and $a\left(c_0\right) = 1.57\pm 0.07$ at 50\,\textdegree C. Both values are in good agreement with the literature.\cite{Landesfeind2019} As explained in Section \ref{model}, the evaluation of the concentration cell data of this system using Newman's concentrated solution theory\cite{Newman2004} deviates from the evaluation using the theory from Latz et al.\cite{Latz2011,Latz2015} by a factor of 2. Together with the VLF-IS measurements determining factor $b\left(0.5\,\mathrm{M}\right)$, we can later use these results to deconvolute $TDF\left(c_0\right)$ and $t_+\left(c_0\right)$ (see Sections \ref{diffusion_coefficient} and \ref{result_VLF-IS}). Further concentration cell measurements with various concentrations and temperatures can be found in the supplementary (see SI Section 2).

\subsection{Galvanostatic Polarization}
\label{diffusion_coefficient}
Using symmetrical Li metal | electrolyte + separator | Li metal cells, we measure the potential response to galvanostatic polarization experiments during the charging $\Delta\Phi\left(t\right)$ and during the relaxation process $\Delta\Phi\left(t^\prime=t-T_\mathrm{I}\right)$ after the current interruption time $T_\mathrm{I}$ (see Section \ref{theo_diff}). For long enough charging times $t$ the concentration gradient within the cell reaches a steady-state, inducing the concentration potential $U_\mathrm{c}$ (see Eq. \ref{U_c}). Specifying $U_\mathrm{c}$ reveals $b\left(c_0\right)$, comprising convoluted information about the transference number $t_+\left(c_0\right)$ and the thermodynamic factor $TDF\left(c_0\right)$. Inserting the previously determined factor $a\left(c_0\right)$ (see Section \ref{sec_conc_cell}) and factor $b\left(c_0\right)$ in Eqs. \ref{decon_t} and \ref{decon_TDF} deconvolutes $t_+\left(c_0\right)$ and $TDF\left(c_0\right)$. The slope of the charging and relaxing potential reveals $D_\pm\left(c_0\right)$ at different time scales. For short times $\Delta\Phi$ shows a linear slope $m_\mathrm{sqrt}$ over $\sqrt{t}$ and $\sqrt{t^\prime}$ respectively. For long times $\Delta\Phi$ exhibits a linear slope $m_\mathrm{ln}$ in a semi-logarithmic plot (see Section \ref{theo_diff}). Inserting the corresponding slopes in Eqs. \ref{D_sqrt} and \ref{D_ln} yields $D_\pm\left(c_0\right)$.\\
Evaluating the potential response $\Delta\Phi$ requires determining the Bruggemann-coefficient $\beta$ of the separator. For this, the authors conduct potentiostatic EIS measurements between every polarization experiment. These measurements reveal the bulk resistance $R_\mathrm{el}$ and the interface resistance $R_\mathrm{int}$ (see SI Section 3.1). Inserting $R_\mathrm{el}$ and the specified conductivity $\kappa\left(c_0\right)$ (see Section \ref{res_cond_cell}) in Eq. \ref{Rel_equ} yields the Bruggemann-coefficient $\beta$.\\
$R_\mathrm{el}$ exhibits almost exclusively small increasing trends at both temperatures (see SI Figure 4a). Evaluating the corresponding mean value yields a Bruggemann-coefficient of $\beta = 2.72\pm 0.09$ and $\beta = 2.5 \pm 0.1$ at 20\,\textdegree C and 50\,\textdegree C respectively. This resembles well the values measured by Landesfeind et al.\cite{Landesfeind2016} for a similar electrolyte at 25\,\textdegree C.\\
While the interface resistance $R_\mathrm{int}$ shows fairly stable behavior at 20\,\textdegree C, it slowly decreases at 50\,\textdegree C (see SI Figure 4b).
\begin{figure*}[tb]%
  \centering
  \begin{subfigure}[t]{0.45\textwidth}
  	\caption{}
	\label{diff_coeff_pot}
	\includegraphics[width=1.0\linewidth]{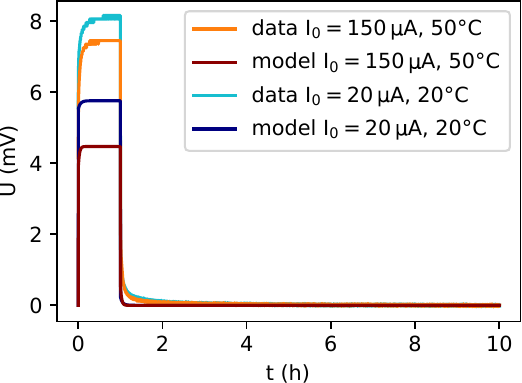}
  \end{subfigure}
  \qquad
  \begin{subfigure}[t]{0.45\textwidth}
  	\caption{}
	\label{diff_coeff_U_c}
  	\includegraphics[width=1.0\linewidth]{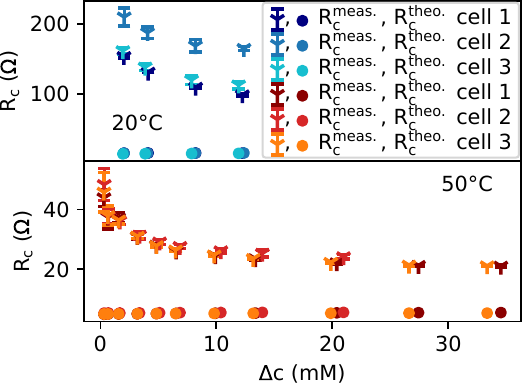}
  \end{subfigure}
  \caption{\textbf{a)} Exemplary polarization measurements with $I_0 = 20$\,\textmu A and $I_0 = 150$\,\textmu A at 20\,\textdegree C and 50\,\textdegree C together with the corresponding simulations. The measurements and the simulations deviate significantly. \textbf{b)} Measured steady-state concentration potentials divided by the corresponding current amplitude $R_\mathrm{c} = \frac{U_\mathrm{c}}{I_0}$ compared to their theoretical expectation. In the experiments, $R_\mathrm{c}$ takes significantly higher values than calculated in the simulations and decreases for increasing concentration gradients $\Delta c$. At 50\,\textdegree C, $R_\mathrm{c}$ converges to a final value.}
\end{figure*}\\
In the galvanostatic polarization experiments, we apply current amplitudes of up to $I_0 = 60$\,\textmu A and $I_0 = 500$\,\textmu A for one hour at 20\,\textdegree C and 50\,\textdegree C (see Table \ref{ac_schedule}). This establishes significant maximum concentration gradients up to $\Delta c = 12\pm 2\,$mM and $\Delta c = 33\pm 6\,$mM. Therefore, the signal-to-noise ratio is enhanced and possible perturbations due to undesired non-diffusive effects are reduced.\\
High current densities could induce non-linear effects. To ensure the validity of our linear response theory we numerically check for non-linear contributions using our 1D-model (see Section \ref{model}, SI Section 3.2). However, the simulation results in negligibly small non-linear contributions, even for the highest applied current amplitudes.\\
In our measurements $\Delta\Phi$ exhibits for all current amplitudes similar behavior resembling the expected shape. Therefore, figure \ref{diff_coeff_pot} exemplary shows one measurement with $I_0 = 20$\,\textmu A at 20\,\textdegree C and one measurement with $I_0 = 150$\,\textmu A at 50\,\textdegree C. In the experiments, the potential relaxes to a small, finite value instead of 0\,V, which we subtract for each data set, respectively. In order to compare the data to the theory, we simulate the polarization experiments for our set-up, combining the electrolyte parameters from Landesfeind et al.\cite{Landesfeind2019} with our 1D-model (see Section \ref{model}).\\
Even though the measurements exhibit the expected shape, they deviate from the simulations. Firstly, the potential reaches higher values during the charging and the relaxation phase. This indicates elevated concentration gradients $\Delta c$ and thus higher concentration potentials $U_\mathrm{c}$ (see Eq. \ref{U_c}) compared to the modeled values. We determine $U_\mathrm{c}$ (see SI Section 3.3) and plot the experimental and theoretical values of $R_\mathrm{c} = \frac{U_\mathrm{c}}{I_0}$ over the corresponding calculated concentration gradient $\Delta c$ in figure \ref{diff_coeff_U_c}. Note, that some experiments using the highest current densities do not yield a reasonable potential response and are therefore omitted. In theory, $R_\mathrm{c}$ stays nearly constant over $\Delta c$ and changes only slightly with $\beta$. In the measurements, $R_\mathrm{c}$ takes significantly higher values than theoretically expected. These findings are similar to the current ratio results from Bergstrom.\cite{Bergstrom2021} Interestingly, the measured $R_\mathrm{c}$ values shrink with increasing current density and thus depend on the induced concentration gradient $\Delta c$. At 50\,\textdegree C, $R_\mathrm{c}$ converges to a final value for the highest concentration gradients. The $I_0 = 10$\,\textmu A reference measurements recorded between the polarization experiments with increasing $I_0$ at 50\,\textdegree C (see Table \ref{ac_schedule}) underline the specific dependence of $R_\mathrm{c}$ on $\Delta c$ (see SI Section 3.3).
\begin{table}[tb]
	\begin{center}
	\caption{Parameters $t_+\left(c_0\right)$ and $TDF\left(c_0\right)$ calculated from $U_\mathrm{c}$. The values deviate from the literature.\cite{Landesfeind2019} Note, that the $TDF\left(c_0\right)$ evaluated by using Newman's theory\cite{Newman2004} deviates for this system by a factor of 2 to our values calculated with the theory from Latz et al.\cite{Latz2011,Latz2015}}\label{tab_diff_t_TDF}
		\begin{tabular}{ccc}	
\toprule		
Parameter & 20\,\textdegree C & 50\,\textdegree C\\
\midrule
$t_+\left(c_0\right)$ & $-8.1$ - $-3.3$ & $-2.8$ - $-0.52$\\
$TDF\left(c_0\right)$ & 0.16 - 0.36 &  0.42 - 1.04\\
\bottomrule	
	\end{tabular}
	\end{center}
\end{table}\\
However, even for the highest current amplitudes $R_\mathrm{c}$ exceeds the expectations. Therefore, calculating factor $b\left(c_0\right)$ using Eq. \ref{U_c} results in a range of elevated values. Inserting factor $a\left(c_0\right)$ and $b\left(c_0\right)$ in Eqs. \ref{factor_a} and \ref{factor_b} deconvolutes $t_+\left(c_0\right)$ and $TDF\left(c_0\right)$. This yields negative values for $t_+\left(c_0\right)$ (see Table \ref{tab_diff_t_TDF}). Even though negative transference numbers have been reported in polymer-based electrolytes\cite{Pesko2017, Pesko2018, Gouverneur2018, Villaluenga2018, Shah2019} we doubt the validity of the negative transference numbers in our system. Further experimental studies using polarization\cite{Landesfeind2019, Nyman2008} and eNMR measurements\cite{Fang2023} as well as molecular dynamics simulations\cite{Bergstrom2021, Ringsby2021, Fang2023} yield positive transference numbers for 1\,M LiPF$_6$ in EC:EMC systems.\\
Secondly, the slopes of the potential $\Delta\Phi$ deviate from the theoretical expectations. Reaching the steady-state during the charging and subsequently relaxing the potential takes a longer duration in the measurements, hinting towards lower diffusion coefficients than reported in the literature.\cite{Landesfeind2019} Evaluating the time evolution of the potential at different time scales specifies $D_\pm\left(c_0\right)$.\\
To examine the short-term behavior we plot $\Delta\Phi^\mathrm{charge}_\mathrm{short} - U_{\bar{i}}$ and $-\Delta\Phi^\mathrm{relax}_\mathrm{short} + U_\mathrm{c}$ over $\sqrt{t}$ and $\sqrt{t^\prime}$ respectively (see Figure \ref{diff_coeff_U_sqrt}). This facilitates the comparison of the charging and relaxation processes since both expressions hold the same theoretical expectation. In the measurements, both processes show almost identical behavior. However, as opposed to the theory, our experiments exhibit solely curved behavior instead of a linear slope at short times. We evaluate the slope $m_\mathrm{sqrt}$ within the time during which the simulations show linear behavior (see SI Section 3.4). The curvature leads to an ambiguous determination of $m_\mathrm{sqrt}$ depending on the specific time interval chosen for the evaluation. Therefore, $D_\pm^\mathrm{sqrt}\left(c_0\right)$ includes a range of possible values. Choosing time intervals containing the shortest times yields the highest diffusion coefficients $D_\pm^\mathrm{sqrt}\left(c_0\right)$ (see Figure \ref{diff_coeff_D_sqrt_short}). Interestingly, $D_\pm^\mathrm{sqrt}\left(c_0\right)$ shows here concentration gradient-dependent behavior. At 50\,\textdegree C, the fastest diffusion coefficients are comparable to the literature.\cite{Landesfeind2019} However, as opposed to the trend reported in the literature $D_\pm^\mathrm{sqrt}\left(c_0\right)$ reaches in our experiments higher values at 20\textdegree C than at 50\,\textdegree C.\\
Thus, the authors doubt that the calculation of $D_\pm^\mathrm{sqrt}\left(c_0\right)$ yields reasonable results. Instead, we believe that the same effect causing the elevated $U_\mathrm{c}$ values also distorts the ratio $\frac{m_\mathrm{sqrt}}{U_\mathrm{c}}$ in Eq. \ref{D_sqrt}.\\
The determination of $D_\pm^\mathrm{ln}\left(c_0\right)$ for long times $t$ and $t^\prime$ does not depend on $U_\mathrm{c}$ (see Eq. \ref{D_ln}). Here, $\ln\left(\Delta \Phi^\mathrm{relax}_\mathrm{long}\right)$ and $\ln\left(\Delta \Phi^\mathrm{charge}_\mathrm{long} - U_{\bar{i}} - U_\mathrm{c}\right)$ theoretically exhibit the same linear slope $m_\mathrm{ln}$ (see Figure \ref{theory_D_log}). In our measurements, the logarithmic terms show indeed similar behavior for the charging and the relaxation process. However, the corresponding slopes are again non-linear until large times (see Figure \ref{diff_coeff_U_log}). For even larger times, the potential resolution of the measurement device limits the visibility of the slope.\\
\begin{figure*}[!t]%
  \centering
  \begin{subfigure}[t]{0.45\textwidth}
        \caption{}
	\label{diff_coeff_U_sqrt}
	\includegraphics[width=1.0\linewidth]{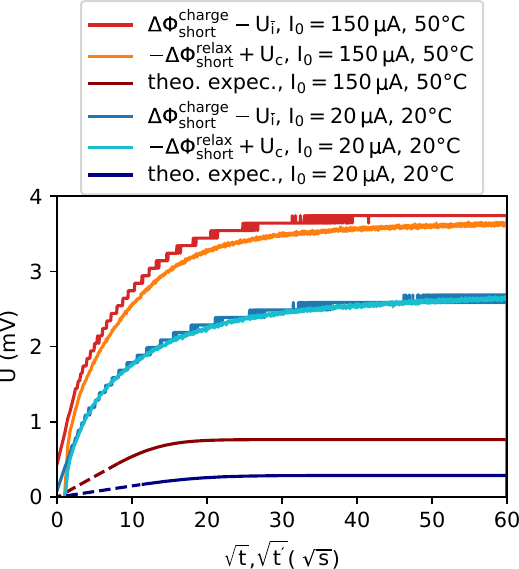}
  \end{subfigure}
  \qquad
  \begin{subfigure}[t]{0.45\textwidth}
        \caption{}
	\label{diff_coeff_U_log}
  	\includegraphics[width=1.0\linewidth]{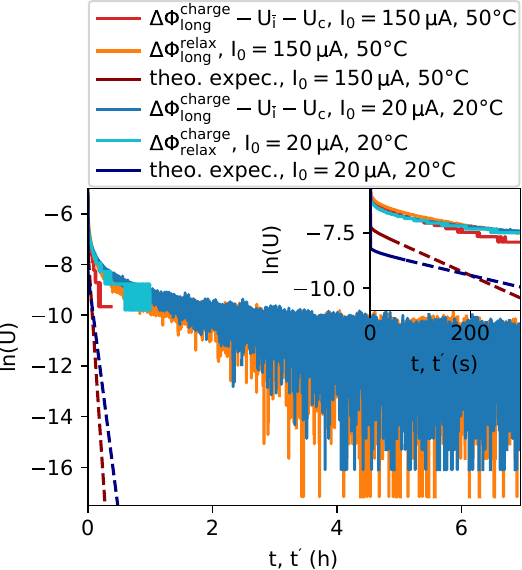}
  \end{subfigure}
  \begin{subfigure}[t]{0.45\textwidth}
        \caption{}
	\label{diff_coeff_D_sqrt_short}
  	\includegraphics[width=1.0\linewidth]{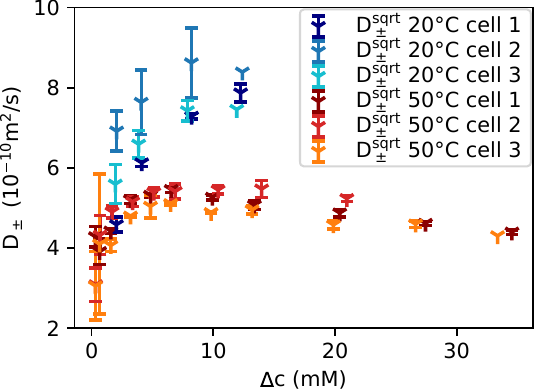}
  \end{subfigure}
  \qquad
  \begin{subfigure}[t]{0.45\textwidth}
        \caption{}
	\label{diff_coeff_D_log}
  	\includegraphics[width=1.0\linewidth]{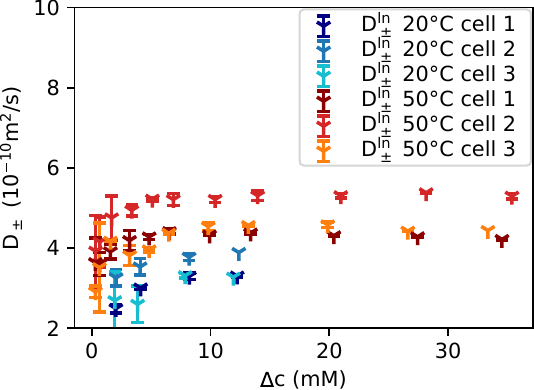}
  \end{subfigure}
  \caption{\textbf{a)} $\Delta\Phi^\mathrm{charge}_\mathrm{short} - U_{\bar{i}}$ and $-\Delta\Phi^\mathrm{relax}_\mathrm{short} + U_\mathrm{c}$ over $\sqrt{t}$ and $\sqrt{t^\prime}$ and the corresponding simulations for $I_0=20$\,\textmu A and $I_0=150$\,\textmu A at 20\,\textdegree C and 50\,\textdegree C. For small times up to $t = t^\prime = 130\,$s and $t = t^\prime = 66\,$s we expect a linear slope (dashed lines). However, the measured potentials show instead curved behavior. \textbf{b)} $\ln\left(\Delta \Phi^\mathrm{relax}_\mathrm{long}\right)$ and $\ln\left(\Delta \Phi^\mathrm{charge}_\mathrm{long} - U_{\bar{i}} - U_\mathrm{c}\right)$ and the corresponding simulations for $I_0 = 20$\,\textmu A and $I_0 = 150$\,\textmu A at 20\,\textdegree C and 50\,\textdegree C in a semi-logarithmic plot. For $t = t^\prime > 70\,$s and $t = t^\prime > 36\,$ we expect a linear slope (dashed lines, see inset). However, the measurements exhibit curved behavior. \textbf{c)} Calculated $D_\pm^\mathrm{sqrt}\left(c_0\right)$ using the time segments which include the shortest times for the fit. For both temperatures $D_\pm^\mathrm{sqrt}\left(c_0\right)$ shows a dependence on the concentration gradient $\Delta c$. \textbf{d)} Calculated $D_\pm^\mathrm{ln}\left(c_0\right)$ using the time segments which include the shortest times for the fit. At both temperatures, $D_\pm^\mathrm{ln}\left(c_0\right)$ increases with increasing $\Delta c$. At 50\,\textdegree C, the coefficient converges to a final value. For c) and d) the error bars represent the respective standard deviation.}
\end{figure*}
We evaluate $m_\mathrm{ln}$ for the time during which we expect linear behavior (see SI Section 3.4). Due to the curvature, $D_\pm^\mathrm{ln}\left(c_0\right)$ comprises again a range of values. For the shortest times during the considered time, $D_\pm^\mathrm{ln}\left(c_0\right)$ takes the highest values (see Figure \ref{diff_coeff_D_log}). The diffusion coefficient also increases with increasing $\Delta c$ and yields for $\Delta c>10\,$mM results comparable to the literature.\cite{Landesfeind2019} However, a unique identification of $D_\pm^\mathrm{ln}\left(c_0\right)$ in our curved data sets is not possible.\\
The polarization experiments show peculiar behavior deviating strongly from the theoretical expectations. Therefore, we doubt that the values for $t_+\left(c_0\right)$, $TDF\left(c_0\right)$ and $D_\pm\left(c_0\right)$ are reliable and suspect additional effects to influence the experiments. The higher concentration potential $U_\mathrm{c}$ hints towards an overall slower, impeded diffusion, leading to higher concentration differences $\Delta c$. Also, the curved behavior of the potential slopes $m_\mathrm{sqrt}$ and $m_\mathrm{ln}$ suggests multiple diffusion processes are present within the cell. Therefore, we suspect further porous structures in addition to the separator.\\
Possible candidates for the porous medium can be found in Talian et al.\cite{Talian2019} The group conducts a combination of polarization and EIS measurements using symmetric Li | separator + electrolyte | Li cells with 1\,M LiTFSI in tetraglyme and and 1,3-dioxolane (1:1, volume). During the polarizations, the pristine Li electrode surface gets covered by islands of mossy Li. For simplicity, the group divides the mossy Li into two porous layers. "Live porous Li" is still electronically connected to the bulk Li and superficially covered with SEI. "Dead porous Li" has lost its electronic connection to the bulk electrode due to passivation. It may even entirely consist of SEI-like products. The coverage of the pristine Li electrode surface with live and dead mossy Li impacts all elements of the impedance data. Note, that ether-based electrolytes exhibit superior stability with respect to Li metal compared to carbonate-based electrolytes.\cite{Liu2023} Therefore, the influence on the impedance data in our study could be even more prominent.\\
The increasing coverage of the electrodes with both porous live and dead Li provides a suitable explanation for the measured trends of the bulk resistance $R_\mathrm{el}$ and the interface resistances $R_\mathrm{int}$ (see SI Section 3.1). Additionally, next to the porous SEI, the growth of mossy Li slows down the ionic transport. The morphology of the porous structure dictates the time constant of the corresponding diffusion processes. A combination of multiple islands with varying morphology on the Li surface has therefore multiple effects. Firstly, the effective diffusion coefficient is lowered, leading to elevated concentration potentials $U_\mathrm{c}$. The steady-state and the ground state are hence reached at much larger time scales. Secondly, the total diffusion process consists of a combination of multiple overlapping diffusion processes with different time constants. This could explain the curved behavior of $m_\mathrm{sqrt}$ and $m_\mathrm{ln}$. Therefore, we cannot isolate the undisturbed diffusion through the separator leading to an ambiguous determination of $D_\pm\left(c_0\right)$ and unreasonable values for $t_+\left(c_0\right)$ and $TDF\left(c_0\right)$.\\
The concentration dependence of $U_\mathrm{c}$ and $D_\pm\left(c_0\right)$ is not quite clear. For low $\Delta c$, the measured potential could interfere with undesired non-diffusive effects. In contrast, elevated concentration gradients enhance the signal-to-noise ratio yielding more accurate results. Furthermore, additional porous structures on the Li surface could impede the diffusion and thus lead to higher $\Delta c$ than theoretically anticipated. Therefore,  non-linear effects and even convection effects could be induced.

\subsection{Very-Low-Frequency Impedance Spectroscopy}
\label{result_VLF-IS}
\begin{figure}[tb]	
	\centering
	\includegraphics[width=1.0\linewidth]{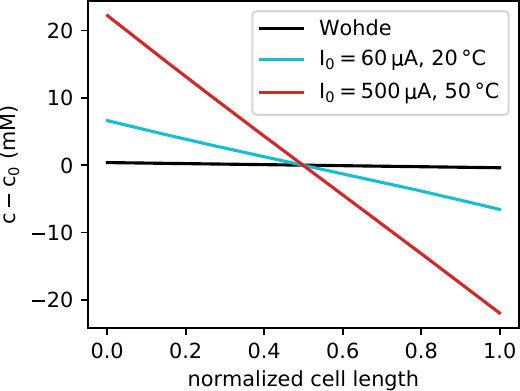}
	\caption{Concentration profiles $c-c_0$ of the largest occurring concentration gradients $\Delta c$ during the VLF-IS measurements. We calculate for the measurement with LP30 conducted by Wohde et al.\cite{Wohde2016} small occurring concentration gradients. In comparison, the VLF-IS measurements with $I_0 = 60$\,\textmu A and $I_0 = 500$\,\textmu A at 20\,\textdegree C and 50\,\textdegree C with our electrolyte yield higher concentration gradients.}
	\label{concentration_profiles}
\end{figure}

In section \ref{sec_conc_cell} concentration cells revealed factor $a\left(c_0\right)$ comprising the transference number $t_+\left(c_0\right)$ and the thermodynamic factor $TDF\left(c_0\right)$ (see Eq. \ref{factor_a}). In section \ref{diffusion_coefficient} measuring the established steady-state concentration potential $U_\mathrm{c}$ in symmetrical Li metal | electrolyte + separator | Li metal cells during galvanostatic polarization deconvoluted both quantities. This gives access to the factor $b\left(c_0\right)$ (see Eq. \ref{U_c}). However, inserting $a\left(c_0\right)$ and $b\left(c_0\right)$ in Eqs. \ref{decon_t} and \ref{decon_TDF} led to negative transference numbers $t_+\left(c_0\right)$. Therefore, the authors suspect that additional diffusion processes through the SEI or mossy Li affect $U_\mathrm{c}$. In this section, we measure EIS down to very low frequencies for three identical symmetrical Li metal | electrolyte + separator | Li metal cells to spectrally isolate the diffusion through the separator from the additional effects. Identifying the corresponding diffusive resistance $R_\mathrm{D}$ allows the determination of the undisturbed factor $b\left(c_0\right)$ (see Eq. \ref{equ_RD}).\\
Wohde et al. have already conducted potentiostatic VLF-IS experiments using symmetrical Li metal | electrolyte | Li metal cells with three different electrolytes.\cite{Wohde2016} However, the group applied very low AC amplitudes of 1-2\,mV$_\mathrm{rms}$. To calculate the corresponding concentration gradients $\Delta c$ we simulate the VLF-IS experiment with our 1D-model (see Section \ref{model}). Taking the electrolyte parameters of LP30 from Landesfeind et al.\cite{Landesfeind2019} and the interface resistance $R_\mathrm{int}$ from Wohde et al.\cite{Wohde2016}, we calculate the occurring maximum concentration difference to be only $\Delta c_\mathrm{max} = 0.8\,$mM during the VLF-IS measurement. Such low concentration gradients could easily interfere with undesired non-diffusive effects, influencing the measured signal. Therefore, the applied current amplitudes in our galvanostatic VLF-IS experiments range from $I_0 = 10$\,\textmu A up to $I_0 = 60$\,\textmu A at 20\,\textdegree C, and from $I_0 = 5$\,\textmu A up to $I_0 = 500$\,\textmu A at 50\,\textdegree C. This corresponds to maximum concentration differences of $\Delta c_\mathrm{max} = 14$\,mM and $\Delta c_\mathrm{max} = 43$\,mM respectively for the highest amplitudes. Figure \ref{concentration_profiles} shows the simulated concentration profiles $c-c_0$ over the normalized cell lengths for the lowest applied frequencies in the corresponding VLF-IS measurements.
\begin{figure}[tb]	
	\centering
	\includegraphics[width=1.0\linewidth]{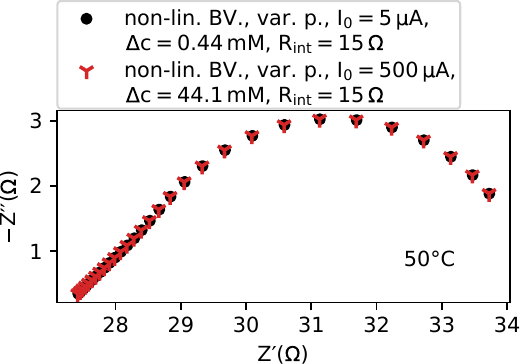}
	\caption{Modeled impedance spectra for $I_0 = 5$\,\textmu A and $I_0 = 500$\,\textmu A. Despite applying a non-linear Butler-Volmer equation and the concentration-dependent electrolyte parameters from Landesfeind et al.,\cite{Landesfeind2019} both spectra show identical results.}
	\label{non_lin_impedance_5_500}
\end{figure}
\begin{figure*}[hbt]%
  \centering
  \begin{subfigure}[t]{\textwidth}
  	\caption{}
	\label{VLF_IS_total_20}
	\includegraphics[width=1.0\linewidth]{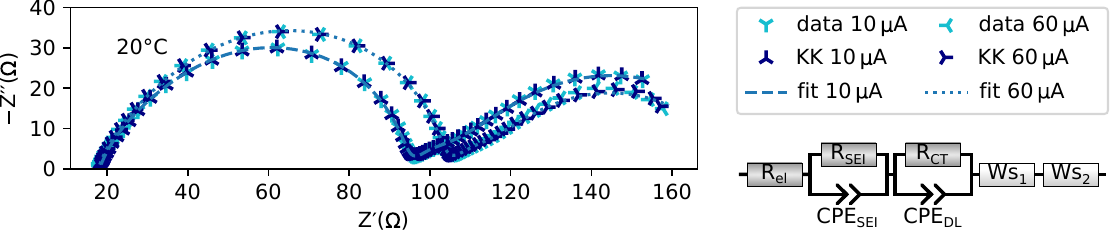}
  \end{subfigure}
  \qquad
  \begin{subfigure}[t]{\textwidth}
  	\caption{}
	\label{VLF_IS_total_50}
  	\includegraphics[width=1.0\linewidth]{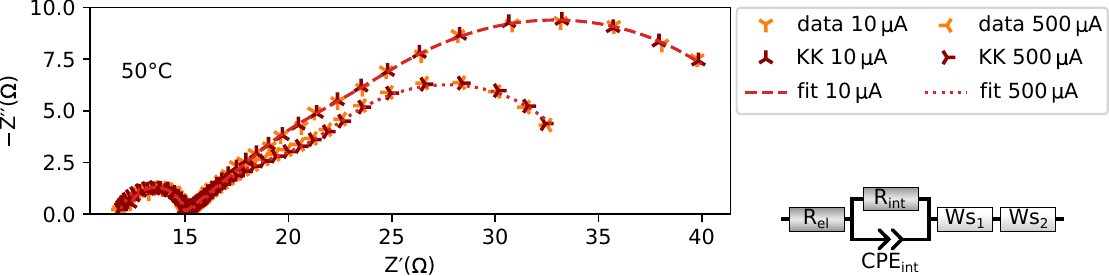}
  \end{subfigure}
  \caption{VLF-IS measurements with amplitudes of 60\,\textmu A, 500\,\textmu A and 10\,\textmu A together with the corresponding equivalent circuits, fits and Kramers-Kronig transformations at 20\,\textdegree C (\textbf{a)}) and 50\,\textdegree C (\textbf{b)}) respectively. The offset of the impedance at high frequencies depicts the bulk electrolyte resistance. The subsequent semi-circle represents interfacial effects, followed by the impedance features due to diffusion.}
  \label{VLF_IS_pic}
\end{figure*}\\
Too high concentration gradients and overpotentials can induce non-linear effects. Hence, we have to ensure that the linearized description of our experiments still applies even for our highest current amplitudes. However, comparing the simulated VLF-IS experiments with $I_0 = 500$\,\textmu A to a linear reference with $I_0 = 5$\,\textmu A at 50\,\textdegree C yields almost no difference. This implies negligibly small non-linear contributions (see Figure \ref{non_lin_impedance_5_500}). Note that in our electro-neutral model, only the concentration gradient induces capacitive behavior. A detailed analysis can be found in the supplementary (see SI Section 4.1). Calculating the Kramers-Kronig transformation of the corresponding experimental data yields an additional linearity check. The transformation and the data exhibit good agreement, deviating even for the highest applied current amplitudes for less than 1\,$\Omega$ (see Figure \ref{residuals}).\\
Figure \ref{VLF_IS_pic} shows exemplary data of our VLF-IS measurements in a Nyquist diagram. For 20\,\textdegree C and 50\,\textdegree C respectively, the plot exhibits one data set with the highest current density $I_0 = 60$\,\textmu A and $I_0 = 500$\,\textmu A and one data set with each $I_0 = 10$\,\textmu A. At 20\,\textdegree C, the cells were still slightly drifting. Therefore, we corrected the VLF-IS data by this deviation on a pro rata temporis basis. For details see SI Section 4.2.\\
The experimental data shows for all measurements a similar form. At high frequencies, the real part of the impedance exhibits an offset. This offset corresponds to the bulk resistance of the electrolyte within the separator. At intermediate frequencies, interface effects induce a subsequent arc. Diffusion effects prevail at the lowest frequencies, forming a second arc.
\begin{figure}[tb]	
	\centering
	\includegraphics[width=1.0\linewidth]{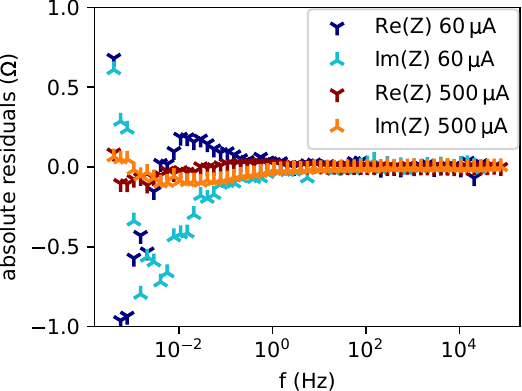}
	\caption{Deviations of the Kramers-Kronig transformation from the measured data at the highest current amplitudes of 60\,\textmu A, 500\,\textmu A at 20\,\textdegree C and 50\,\textdegree C respectively. The residuals are for all measurements lower than 1\,$\Omega$.}
	\label{residuals}
\end{figure}\\
Using the RelaxIS software, we represent the data with the equivalent circuits shown in figure \ref{VLF_IS_pic}. A serial resistance corresponds to the bulk electrolyte resistance $R_\mathrm{el}$. For the measurements at 20\,\textdegree C, two parallel R-CPE elements account for the interfacial impedance $Z_\mathrm{int} = Z_\mathrm{SEI} + Z_\mathrm{CT}$ (see Figure \ref{VLF_IS_total_20}). These incorporate the SEI impedance, charge transfer and double layer processes. At 50\,\textdegree C, one single parallel R-CPE element summarising these effects sufficiently represents the first arc (see Figure \ref{VLF_IS_total_50}). In contrast to the theory, the data exhibits at least two features at 20\,\textdegree C at low frequencies. These features are also visible for higher current amplitudes at 50\,\textdegree C. Therefore, the equivalent circuits include two serial Warburg short elements (Ws) in order to fit the diffusive behavior.\\
Determining the bulk resistance $R_\mathrm{el}$ reveals the Bruggemann-coefficient $\beta$ of the separator (see Eq. \ref{Rel_equ}), required for the evaluation of the resonance frequency and factor $b\left(c_0\right)$ (see SI Section 4.3). At both temperatures, $R_\mathrm{el}$ exhibits a slow upward trend for all cells. Only one cell shows a downward trend at 20\,\textdegree C. Together with the previously determined conductivities (see Section \ref{res_cond_cell}), averaging all results for $R_\mathrm{el}$ at 20\,\textdegree C and at 50\,\textdegree C yields the Bruggemann-coefficients $\beta = 3.1\pm 0.3$ and $\beta = 2.9\pm 0.3$ respectively. These values are slightly higher than the coefficients in the polarization experiments in section \ref{diffusion_coefficient}.\\
Tracking the interface resistance $R_\mathrm{int}$ allows detecting changes to the interface between the Li electrode and the separator. At 20\,\textdegree C, $R_\mathrm{int}$ increases linearly over time (see SI Figure 14b). We correct this drift as mentioned above. At 50\,\textdegree C, the resistance decreases.
\begin{figure*}[tb]%
  \centering
  \begin{subfigure}[t]{0.45\textwidth}
  	\caption{}
	\label{Ws1}
	\includegraphics[width=1.0\linewidth]{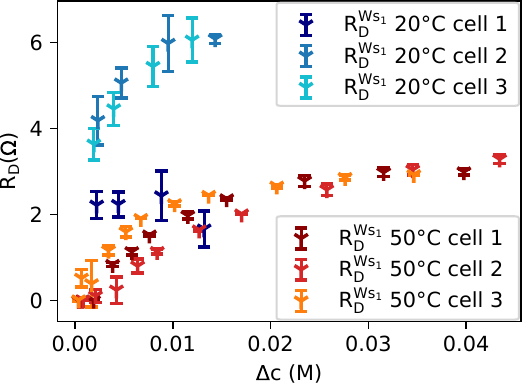}
  \end{subfigure}
  \qquad
  \begin{subfigure}[t]{0.45\textwidth}
  	\caption{}
	\label{Ws2}
  	\includegraphics[width=1.0\linewidth]{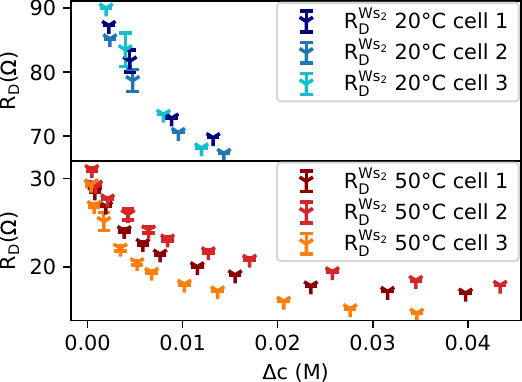}
  \end{subfigure}
  \caption{\textbf{a)} The resistance of Ws$_1$ rises with increasing concentration gradient and converges to a final value at 50\,\textdegree C. The increase could originate from a growing solid-electrolyte interphase (SEI) or mossy Li. \textbf{b)} The resistance of Ws$_2$ decreases with increasing concentration gradient and converges to a final value at 50\,\textdegree C. The origin of the decrease is not quite clear. The error bars represent the respective standard deviation.}
\end{figure*}\\
The diffusive impedance should in theory induce one single arc at low frequencies. The amplitude $R_\mathrm{D}$ of this arc comprises the desired factor $b\left(c_0\right)$ (see Eq. \ref{equ_RD}). However, our measurements exhibit at least two features Ws$_1$ and Ws$_2$. Both could correspond to the diffusion through the separator. Therefore, we calculate $b\left(c_0\right)$ for Ws$_1$ and Ws$_2$.\\
Ws$_1$ and Ws$_2$ strongly overlap. This allows identifying Ws$_1$ only at 20\,\textdegree C and for high current densities at 50\,\textdegree C. At low current densities, using one single Ws element for the total diffusive resistance yields the best fits.\\
The shape of Ws$_1$ and Ws$_2$ deviates from the theoretical expectation. Both features show depressed arcs with lower inclination angles than 45\,\textdegree. This could hint towards a pore morphology within the separator or possible porous structures on the Li electrode surface\cite{Loew2023} which our 1D model can not accurately describe. Therefore, Eq. \ref{equ_RD} yields unreasonable fits. Using the modified equation Eq. \ref{Warburg_modified} with $\alpha \neq 0.5$ fits the measured data fairly well. This expression assumes, that the amplitude of the arc $R_\mathrm{D}$ still matches the theoretical expression but the capacitive behavior is modified. Therefore, also the corresponding resonance frequency shifts. However, due to this discrepancy, the following analysis of the fitting results with $\alpha \neq 0.5$ using our theory is questionable.\\
Ws$_1$ resonates around $f = 79\pm 31$\,mHz at 20\,\textdegree C and $f = 27\pm 9$\,mHz at 50\,\textdegree C, exceeding their theoretical expectations of $f_{\alpha = 0.5} = 0.7\pm 0.2$\,mHz and $f_{\alpha = 0.5} = 1.3\pm 0.3$\,mHz. The measured amplitude $R_\mathrm{D}^\mathrm{Ws_1}$ rises with increasing concentration gradient at both temperatures (see Figure \ref{Ws1}) but converges to a final value at 50\,\textdegree C. Inserting $R_\mathrm{D}^\mathrm{Ws_1}$ and the previously determined Bruggemann-coefficient $\beta$ in Eq. \ref{equ_RD} yields factor $b\left(c_0\right)$. Since $R_\mathrm{D}^\mathrm{Ws_1}$ depends on the concentration gradient, $b\left(c_0\right)$ comprises a range of values. Deconvoluting factor $a\left(c_0\right)$ (see Section \ref{sec_conc_cell}) and factor $b\left(c_0\right)$ with Eqs. \ref{decon_t} and \ref{decon_TDF} isolates the transference number $t_+\left(c_0\right)$ and the thermodynamic factor $TDF\left(c_0\right)$. The corresponding results are shown in Table \ref{tab_t_TDF}. Compared to the literature\cite{Landesfeind2019} we obtain elevated values for $t_+\left(c_0\right)$ and $TDF\left(c_0\right)$. Therefore, Ws$_1$ does not match the expected impedance for the diffusion through the separator neither in resonance nor in amplitude.\\
\begin{table}[tb]
	\begin{center}
	\caption{Parameters $t_+\left(c_0\right)$ and $TDF\left(c_0\right)$ calculated from the diffusion features Ws$_1$ and Ws$_2$. The values deviate from the literature.\cite{Landesfeind2019} Note, that the $TDF\left(c_0\right)$ evaluated by using Newman's theory\cite{Newman2004} deviates for this system by a factor of 2 to our values calculated with the theory from Latz et al.\cite{Latz2011,Latz2015}}\label{tab_t_TDF}
		\begin{tabular}{cccc}	
\toprule		
Feature	& Parameter & 20\,\textdegree C	& 50\,\textdegree C \\
\midrule
\multirow{2}{*}{Ws$_1$} & $t_+\left(c_0\right)$ & 0.75 - 0.94 & 0.78 - 1.0\\
	& $TDF\left(c_0\right)$ & 6.2 - 23.8 &  7.4 - Inf\\
 & & &\\
 \multirow{2}{*}{Ws$_2$} & $t_+\left(c_0\right)$ &  $-2.6$ - $-1.2$ & $-1.0$ - $-0.01$\\
	& $TDF\left(c_0\right)$ & 0.42 - 0.70 & 0.78 - 1.56\\
\bottomrule	
	\end{tabular}
	\end{center}
\end{table}
Ws$_2$ has a lower resonance frequency of $f = 0.8\pm 0.2$\,mHz at 20\,\textdegree C and $f = 1.2\pm 0.2$\,mHz at 50\,\textdegree C. This is in better agreement with the theoretical expectation than Ws$_1$. However, also Ws$_2$ exhibits $\alpha \neq 0.5$. Opposed to $R_\mathrm{D}^\mathrm{Ws_1}$ the amplitude $R_\mathrm{D}^\mathrm{Ws_2}$ decreases for increasing concentration gradients, but also converges to a final value at 50\,\textdegree C (see Figure \ref{Ws2}). Calculating $b\left(c_0\right)$ and deconvoluting $t_+\left(c_0\right)$ and $TDF\left(c_0\right)$ yields thoroughly negative values for the transference number (see Table \ref{tab_t_TDF}). Therefore, the authors suspect that Ws$_2$ does not represent the sole diffusion through the separator but comprises additional, overlapping diffusion effects. This makes a reasonable evaluation of factor $b\left(c_0\right)$ unfeasible.\\
As mentioned in section \ref{diffusion_coefficient}, layers of live and dead mossy Li could influence the impedance data.\cite{Talian2019} The additional porous layers affect the bulk resistance $R_\mathrm{el}$ and the interfacial resistance $R_\mathrm{int}$ (see SI Section 4.3). Next to the diffusion through the porous SEI, the diffusion through mossy Li also impedes the ionic transport. Therefore, additional features and a higher total diffusive resistance occur in the impedance data at low frequencies. Depending on the corresponding morphology these features may overlap with the diffusion through the separator, and thus, be hard to distinguish. Evaluating the diffusive resistance without clear identification of the occurring processes leads to erroneous transference numbers.\\
Our measurements exhibit two overlapping features at low frequencies, Ws$_1$ and Ws$_2$. However, analyzing Ws$_1$ yields a higher resonance frequency and a lower amplitude as expected for the diffusion through the separator. Therefore, Ws$_1$ could originate from diffusion through the SEI or mossy Li.\\
Ws$_2$ resonates fairly close to the expected resonance frequency. However, the corresponding amplitudes are very high, leading to negative transference numbers. The authors therefore suspect additional diffusive processes through mossy Li layers overlapping with Ws$_2$. \\
We try to isolate the diffusion through the separator by calculating the distribution of relaxation times (DRT)\cite{Schichlein2002, IVERSTIFF&EacuteE2017, Danzer2019} (see SI Section 4.5). The DRT transforms the impedance data into a distribution of time constants, corresponding to several serial RC-elements. This allows resolving processes with similar resonance frequencies with higher resolution. However, even using the DRT reveals only two distinguishable diffusive features, corresponding to Ws$_1$ and Ws$_2$.\\
The porous mossy Li layers on the pristine Li electrode seem to explain most of the phenomena. However, these layers have to grow within the first conditioning cycle at 20\,\textdegree C (see SI Section 4.6). Rapidly changing the sign of the applied current during the VLF-IS conditioning may also lead to a porous mix of SEI and Li instead of the more structured layers described by Talian et al.\cite{Talian2019}\\
Ws$_1$ and Ws$_2$ show concentration gradient-dependent behavior. The decrease of $R_\mathrm{D}^\mathrm{Ws_2}$ overcompensates the increase of $R_\mathrm{D}^\mathrm{Ws_1}$. Therefore, increasing the concentration gradient leads to a significant decrease in the total diffusive resistance $R_\mathrm{D}^\mathrm{tot} = R_\mathrm{D}^\mathrm{Ws_1} + R_\mathrm{D}^\mathrm{Ws_2}$. Thereby, the $I_0 = 10$\,\textmu A reference impedance experiments verify the dependence of $R_\mathrm{D}^\mathrm{tot}$ on $\Delta c$ (see SI Section 4.4). This is consistent with the behavior of $R_\mathrm{c}$ in section \ref{diffusion_coefficient}.\\
This concentration gradient dependence is not quite clear and may have multiple contributions. Firstly, Ws$_1$ and Ws$_2$ are more distinguishable at elevated current amplitudes. Therefore, the fitting process may assign Ws$_1$ higher and Ws$_2$ lower resistance values $R_\mathrm{D}$ with increasing concentration gradients. Secondly, Ws$_1$ probably represents the diffusion through the SEI or mossy Li. Therefore, this feature might grow over time. However, this alone does not explain the decrease of the total diffusive resistance $R_\mathrm{D}^\mathrm{tot} = R_\mathrm{D}^\mathrm{Ws_1} + R_\mathrm{D}^\mathrm{Ws_2}$ with increasing concentration gradients. Thirdly, the diffusive effects could be perturbed by undesired non-diffusive effects at low concentration gradients. Therefore, the signal-to-noise ratio might be enhanced at elevated concentration gradients. Fourthly, additional layers of live or dead mossy Li impede the diffusion processes and may lead to higher concentration gradients than considered in our model calculations. Therefore, we can not ensure the absence of non-linear effects. Very high concentration gradients might even induce convection effects within the separator.\\

\section{Conclusion}
In this work, we determine the four electrolyte parameters conductivity $\kappa\left(c_0\right)$, diffusion coefficient $D_\pm\left(c_0\right)$, transference number $t_+\left(c_0\right)$ and thermodynamic factor $TDF\left(c_0\right)$ for 0.5\,M LiPF$_6$ in EC:EMC (3:7, weight) at 20\,\textdegree C and 50\,\textdegree C, using concentration cell measurements, galvanostatic pulse polarization experiments and EIS.\\
We measure the conductivity of the sole electrolyte using the airtight TSC 1600 closed cell (rhd instruments). Applying potentiostatic EIS from 8\,MHz to 5\,kHz with an amplitude of 30\,mV reveals $\kappa\left(c_0\right) = 7.1\pm 0.2\,\frac{\mathrm{mS}}{\mathrm{cm}}$ at 20\,\textdegree C and $\kappa\left(c_0\right) = 9.6\pm 0.3\,\frac{\mathrm{mS}}{\mathrm{cm}}$ at 50\,\textdegree C. These values are in good agreement with the literature.\\
Using the concentration cells we measure the potential between two Li electrodes immersed in 0.25\,M and 0.75\,M LiPF$_6$ in EC:EMC (3:7, weight). The potential comprises a convoluted expression of $t_+\left(c_0\right)$ and $TDF\left(c_0\right)$ in terms of factor $a\left(c_0\right) = TDF\left(c_0\right)(1-t_+\left(c_0\right))$. In our experiments, $a\left(c_0\right)$ yields similar values as reported in the literature\cite{Landesfeind2019} with $a\left(c_0\right) = 1.54\pm 0.06$ at 20\,\textdegree C and $a\left(c_0\right) = 1.57\pm 0.07$ at 50\,\textdegree C.\\
To deconvolute $t_+\left(c_0\right)$ and $TDF\left(c_0\right)$ we apply various galvanostatic, 1\,h long pulses to symmetric Li metal | electrolyte + separator | Li metal cells. The amplitudes range from $\pm 10$\,\textmu A to $\pm 60$\,\textmu A at 20\,\textdegree C and from $\pm 5$\,\textmu A to $\pm 500$\,\textmu A at 50\,\textdegree C to induce significant concentration gradients. The established steady-state concentration potential response $U_\mathrm{c}$ during the pulses allows the determination of factor $b\left(c_0\right)$. $b\left(c_0\right)$ contains a slightly different convoluted expression of $t_+\left(c_0\right)$ and $TDF\left(c_0\right)$ than factor $a\left(c_0\right)$. Combining $a\left(c_0\right)$ and $b\left(c_0\right)$ isolates $t_+\left(c_0\right)$ and $TDF\left(c_0\right)$.  However, in our polarization experiments, $U_\mathrm{c}$ reaches higher values than theoretically expected. Therefore, also factor $b\left(c_0\right)$ yields elevated values. This results in negative transference numbers $t_+\left(c_0\right)$ for both temperatures, resembling the findings of Bergstrom et al.\cite{Bergstrom2021} We agree with their conclusion, that the negative transference numbers measured in this manner do not represent the true quantity of the electrolyte system.\\
The slope of the potential $\Delta\Phi$ during the charging and relaxation process of the polarization experiments theoretically reveals the diffusion coefficient $D_\pm\left(c_0\right)$. For short times $\Delta\Phi$ shows a linear slope over $\sqrt{t}$ and $\sqrt{t^\prime}$ respectively. For long times $\Delta\Phi$ exhibits a linear slope in a semi-logarithmic plot. However, in our experiments, the slope does not show the expected linear behavior in either data-representation. Therefore, the determination of the linear slope and the corresponding calculation of $D_\pm\left(c_0\right)$ is ambiguous, depending on the chosen time interval for the fit. This makes the unique identification of $D_\pm\left(c_0\right)$ challenging, especially for electrolytes with unknown diffusion coefficients.\\
The elevated concentration potential and the ambiguous diffusion coefficient suggest additional porous structures on the Li metal electrodes impeding the diffusion processes similar to the findings of Talian et al.\cite{Talian2019} Therefore, $R_\mathrm{c}$ and $D_\pm\left(c_0\right)$ do not only capture the diffusion through the separator but also the diffusion through the additional structures.\\
In order to spectrally isolate the low-frequency diffusion through the separator we conduct galvanostatic VLF-IS from 4\,MHz down to 400\,\textmu Hz with AC-amplitudes from 10\,\textmu A to 60\,\textmu A at 20\,\textdegree C and from 5\,\textmu A to 500\,\textmu A. The impedance data as well as the corresponding DRT show in fact at least two Warburg-short-like, distinguishable features at low frequencies. However, the amplitudes of both features deviate from the expected value, yielding either very high or again negative transference numbers. Therefore, the authors suspect that the feature with higher resistance comprises the diffusion through the separator but is still influenced by the additional porous structures. Hence, even the VLF-IS measurements do not yield feasible results for identifying $t_+\left(c_0\right)$ and $TDF\left(c_0\right)$.\\
In summary, additional porous structures on the Li metal electrodes make an explicit determination of the transference number and the thermodynamic factor for our electrolyte system difficult. Therefore, the development of non-blocking electrodes avoiding these additional structures could lead to more accurate results.\cite{Chae2023} Alternatively, in-operando NMR or eNMR spectroscopy\cite{Lorenz2022,Kilchert2023,Gouverneur2015,Rosenwinkel2019, Steinrueck2020} operate as suitable methods for measuring the desired quantities.

\clearpage

\section*{Acknowledgements}

This work was supported by the project DeepBat, a collaboration between National Aeronautics and Space Administration – Jet Propulsion Laboratory (NASA - JPL) and the German Aerospace Center (DLR). The authors express their gratitude to the Center for Electrochemical Energy Storage Ulm \& Karlsruhe (CELEST) and thank Dr. Marcel Drüschler (rhd instruments) for fruitful discussions.

\section*{Conflict of Interest}

The authors declare no conflict of interest.

\begin{shaded}
\noindent\textsf{\textbf{Keywords:} \keywords} 
\end{shaded}


\setlength{\bibsep}{0.0cm}
\bibliographystyle{Wiley-chemistry}
\bibliography{example_refs}

\end{document}